\documentclass[aps,prd,reprint,nofootinbib,preprintnumbers,eqsecnum]{revtex4-1}
\usepackage{newtxtext}
\usepackage[varg]{newtxmath}
\usepackage[T1]{fontenc}
\usepackage{graphicx}

\usepackage{titlesec}
\titleformat{\paragraph}[runin]{\itshape}{}{0em}{}[.---]
\titlespacing{\paragraph}{\the\parindent}{0em}{0em}

\pdfoutput=1
\usepackage{microtype}
\usepackage[all]{nowidow}
\let\revtitle\maketitle
\renewcommand{\maketitle}{%
	\revtitle
	\tolerance=200
	\hyphenpenalty=200
}

\renewcommand{\dot}[1]{\overset{\bm .}{#1}\vphantom{#1}}

\DeclareFontFamily{U}{futm}{}
\DeclareFontShape{U}{futm}{m}{n}{<-> fourier-bb}{}
\DeclareMathAlphabet{\mathbb}{U}{futm}{m}{n}

\DeclareSymbolFont{cmreg}{OT1}{cmr}{m}{n}
\DeclareSymbolFont{cmmath}{OML}{cmm}{m}{i}
\DeclareSymbolFont{cmsymbols}{OMS}{cmsy}{m}{n}
\DeclareSymbolFont{cmlargesymbols}{OMX}{cmex}{m}{n}
\DeclareSymbolFontAlphabet{\mathcal}{cmsymbols}

\DeclareMathSymbol{\partial}{0}{cmmath}{64}
\DeclareMathSymbol{g}{\mathalpha}{cmmath}{103}

\DeclareMathSymbol{\ointop}{\mathop}{cmlargesymbols}{72}
\DeclareMathSymbol{\intop}{\mathop}{cmlargesymbols}{82}
\DeclareMathDelimiter{(}{\mathopen}{cmreg}{40}{cmlargesymbols}{0}
\DeclareMathDelimiter{)}{\mathclose}{cmreg}{41}{cmlargesymbols}{1}

\usepackage{titlesec}
\newcommand{\mysectionnumbering}{\thesection.~}

\titleformat{\section}{\bfseries\center\uppercase}{\mysectionnumbering}{0em}{}
\titleformat{\subsection}{\bfseries\center}{}{0.1em}{\thesubsection.~}
\titleformat{\subsubsection}{\bfseries\itshape\center}{}{0.1em}{\thesubsubsection.~}

\titlespacing{\section}{0pt}{1.7em plus 0.1em minus 0.1em}{0.7em}
\titlespacing{\subsection}{0pt}{1.5em}{0.5em}
\titlespacing{\subsubsection}{0pt}{1.5em}{0.5em}

\titleformat{\paragraph}[runin]{\itshape}{}{0em}{}[.---]
\titlespacing{\paragraph}{\the\parindent}{0em}{0em}

\let\oldappendix\appendix

\usepackage{etoolbox}
\makeatletter
\renewcommand{\appendix}{\@ifstar\oneappendix\manyappendices}
\makeatother

\newcommand{\defappsec}{%
	\let\oldsection\section
	\renewcommand{\section}[1]{%
	\ifstrempty{##1}
		{\oldsection{}}
		{\oldsection{:~{##1}}}
	}
}

\newcommand{\oneappendix}{%
\oldappendix*
\defappsec
\renewcommand{\mysectionnumbering}{\MakeUppercase{Appendix}}
\renewcommand\theequation{\Alph{section}\arabic{equation}}
}
\newcommand{\manyappendices}{%
\oldappendix
\defappsec
\renewcommand{\mysectionnumbering}{\MakeUppercase{Appendix}~\thesection}
\renewcommand\theequation{\Alph{section}\arabic{equation}}
}

\makeatletter
\renewcommand\@makefntext[1]{%
\noindent{\hspace{1em}}{\@makefnmark}#1}
\makeatother

\makeatletter
\renewcommand\@makefnmark{\hbox{\color{black}\@textsuperscript{\normalfont\@thefnmark}}}
\makeatother

\renewcommand{\footnoterule}{%
  \kern -3pt
  \hrule width 1.2cm
  \kern 4pt
}

\usepackage[dvipsnames]{xcolor}
\definecolor{revblue}{HTML}{2d3092}
\colorlet{blue}{revblue}

\let\revcite\cite
\renewcommand\cite[1]{{\color{blue}\revcite{#1}}}

\let\reveqref\eqref
\renewcommand\eqref[1]{{\color{blue}\reveqref{#1}}}

\usepackage{xparse}
\ExplSyntaxOn
\DeclareExpandableDocumentCommand{\IfNoValueOrEmptyTF}{mmm}{%
	\IfNoValueTF{#1}{#2}{%
    	\tl_if_empty:nTF {#1} {#2} {#3}
    }
}
\DeclareDocumentCommand{\IfNoValueOrEmptyF}{mm}{%
	\IfNoValueOrEmptyTF{#1}{}{#2}
}
\ExplSyntaxOff

\newlength{\dskip}
\DeclareDocumentCommand{\AddDisplaySkip}{oo}{%
	\IfNoValueOrEmptyF{#1}{%
		\addtolength\abovedisplayskip{#1\dskip}
		\addtolength\abovedisplayshortskip{#1\dskip}
	}
	\IfNoValueOrEmptyF{#2}{%
		\addtolength\belowdisplayskip{#2\dskip}
		\addtolength\belowdisplayshortskip{#2\dskip}
	}
}
\DeclareDocumentCommand{\SetDisplaySkip}{oo}{%
	\IfNoValueOrEmptyF{#1}{%
		\setlength\abovedisplayskip{#1\dskip}
		\setlength\abovedisplayshortskip{#1\dskip}
	}
	\IfNoValueOrEmptyF{#2}{%
		\setlength\belowdisplayskip{#2\dskip}
		\setlength\belowdisplayshortskip{#2\dskip}
	}
}

\usepackage[
	hyperfootnotes=false,
	colorlinks=true,
	allcolors=blue,
	breaklinks=true]
{hyperref}

\usepackage{bm}
\newcommand{\dx}{\text{d}}
\newcommand{\bmf}[1]{\mathbf{#1}}
\newcommand{\avg}[1]{{\langle #1 \rangle}}
\newcommand{\w}{\omega}
\let\Re\relax
\DeclareMathOperator{\Re}{Re}

\newcommand{\Y}{\mathcal{Y}}
\newcommand{\bs}{u}
\newcommand{\cs}{w}

\begin{document}
\title{Evolution of diffuse scalar clouds around binary black holes}

\author{Leong Khim \surname{Wong}}
\email{L.K.Wong@damtp.cam.ac.uk}
\affiliation{DAMTP, Centre for Mathematical Sciences, University of Cambridge,\protect\\
Wilberforce Road, Cambridge CB3 0WA, United Kingdom}

\begin{abstract}
The use of modern effective field theory techniques has sparked significant developments in many areas of physics, including the study of gravity. Case in point, such techniques have recently been used to show that binary black holes can amplify incident, low-frequency radiation due to an interplay between absorption at the horizons and momentum transfer in the bulk of the spacetime. In this paper, we further examine the consequences of this superradiant mechanism on the dynamics of an ambient scalar field by taking the binary's long-range gravitational potential into account at the nonperturbative level. Doing so allows us to capture the formation of scalar clouds that are gravitationally bound to the binary. If the scalar is light enough, the cloud can be sufficiently diffuse (i.e., dilute while having considerable spatial extent) that it engulfs the binary as a whole. Its subsequent evolution exhibits an immensely rich phenomenology, which includes exponential growth, beating patterns, and the upscattering of bound states into scalar waves. While we find that these effects have negligible influence on the binary's inspiral in the regime wherein our approximations are valid, they offer new, analytic insight into how binary black holes interact with external perturbations. They may also provide useful, qualitative intuition for interpreting the results from future numerical simulations of these complex~systems.
\end{abstract}

\maketitle

\section{Introduction}

The details of how binary black holes evolve and coalesce are now well established \cite{Blanchet:2013haa, Centrella:2010zf, Lehner:2014asa, Cardoso:2014uka, Damour:2012mv, Kokkotas:1999bd, Sasaki:2003xr, Poisson:2011nh}. Indeed, our ability to extract gravitational-wave signals from data gathered by the LIGO Scientific and Virgo collaborations \cite{Abbott:2016blz, TheLIGOScientific:2016pea,*TheLIGOScientific:2016pea_E, LIGOScientific:2018mvr} is a testament to how accurately we understand these systems according to general relativity. That being said, while we now have a strong grasp of how binary black holes behave when isolated in empty space, their dynamical response to external perturbations is less well~understood.

The presence of some matter distribution on top of this spacetime generically introduces multiple new scales into the problem, which can lead to a myriad of new effects. As a first step, recent work in addressing this question has focused on the ramifications of perturbing a binary black hole with a Klein--Gordon field.%
\footnote{In a related context, a significant amount of work has already been undertaken to quantify the astrophysical systematics introduced by accretion disks, magnetic fields, and tidal perturbations \cite{Barausse:2014tra, Barausse:2014pra, Palenzuela:2010nf, Giacomazzo:2012iv, Yunes:2011ws, Kocsis:2011dr, Kocsis:2012cs, Kocsis:2012ui, Derdzinski:2018qzv,*Derdzinski:2018qzv_E, Barausse:2007dy, Bonga:2019ycj}.}
While a scalar field is a natural starting point due to its technical simplicity, it is also of particular interest on phenomenological grounds, as the existence of new ultralight fundamental fields is a generic prediction of string theory \cite{Svrcek:2006yi, Arvanitaki:2009fg, Jaeckel:2010ni, Marsh:2015xka, Hui:2016ltb}. Because these hypothetical particles may couple only very weakly to the Standard Model, prospects for their detection rest on finding novel and innovative probes. In this regard, the advent of gravitational-wave astronomy is particularly opportune \cite{Barack:2018yly, Sathyaprakash:2019yqt, Bertone:2019irm, Barausse:2020rsu}.

On one end of the spectrum, it is now known that ultralight bosons can rapidly form ``clouds'' around a rotating black hole due to a superradiant instability, which is most pronounced when the field's Compton wavelength is comparable to the gravitational radius of the hole
\cite{Bekenstein:1998nt, Brito:2015oca, Detweiler:1980uk, Cardoso:2004nk,*Cardoso:2004nk_E, Dolan:2007mj, Dolan:2012yt, Dolan:2018dqv, Pani:2012vp, Pani:2012bp, Witek:2012tr, Brito:2014wla, Arvanitaki:2010sy, Yoshino:2013ofa, Zilhao:2015tya, Baryakhtar:2017ngi, East:2017ovw, East:2017mrj, East:2018glu, Brito:2017wnc,*Brito:2017zvb, Hannuksela:2018izj, Cardoso:2018tly, Brito:2013wya, Brito:2020lup, Herdeiro:2014goa, Herdeiro:2016tmi, Herdeiro:2017phl, Degollado:2018ypf}.
If such a cloud forms around a black hole belonging to a binary system, multiple studies~\cite{Baumann:2018vus, Baumann:2019eav, Baumann:2019ztm, Berti:2019wnn, Zhang:2018kib, Zhang:2019eid} have shown that this scalar (or vector) cloud can lead to a large dephasing of the gravitational-wave signal. A key prediction is that these clouds can undergo resonant transitions during the early inspiral phase, when the cloud's characteristic size $\lambda$ is much smaller than the binary's orbital separation $a$.
For field configurations with even larger spatial extent, Bernard~et~al.~\cite{Bernard:2019nkv} recently reported the existence of global quasinormal modes, which were observed in numerical simulations of a scalar Gaussian pulse of size ${\lambda \sim a}$ scattering off a binary. Last, working analytically in the long-wavelength limit (${\lambda \gg a}$), we recently demonstrated that binary black holes can amplify incident, low-frequency radiation under certain conditions~\cite{Wong:2019kru}.

These advances notwithstanding, there is still much to learn about how binary black holes interact with external fields and whether these interactions can be used to search for new physics. The goal of this present paper is to further our understanding of this problem in the long-wavelength limit by incorporating the effects of the binary's long-range gravitational potential at the nonperturbative level. In doing so, we will be able to study not just the scattering of radiation but also the formation and evolution of scalar clouds diffuse enough to engulf the binary as a whole. As we will show, the same mechanism that led to the amplification of radiation discovered in Ref.~\cite{Wong:2019kru} triggers, among other things, the exponential growth or decay of these~clouds.

At the heart of this phenomenon is the absorptive nature of the binary's constituents:~Because an ambient field must always satisfy purely ingoing boundary conditions at the horizons, it is ``dragged'' alongside the black holes as~they orbit one another.~Consequently, while a part of this scalar field irreversibly crosses the horizons and deposits its energy into the individual black holes, what remains in the bulk spacetime is agitated by the binary's motion and can either gain or lose momentum as a result.~Together, these two effects facilitate a direct exchange of energy and momentum between the binary and the scalar field, which can proceed in either direction depending on their relative phase velocities.~Given this description, the close parallel between this process and the reflection of electromagnetic waves off moving conductive boundaries \cite{VanBladel1976, Cooper1980, Cooper1993} is unsurprising. The underlying mathematics also shares many features in common with superradiant phenomena in general \cite{Bekenstein:1998nt, Brito:2015oca, Richartz:2009mi}, making it natural to regard this mechanism as a novel variant of superradiance fueled by the binary's orbital motion. Accordingly, we will refer to this mechanism as \emph{orbital superradiance}.

The remainder of this paper proceeds as follows. In Sec.~\ref{sec:eft}, we begin by reviewing the low-energy effective field theory (EFT) constructed in Ref.~\cite{Wong:2019kru} to describe the propagation of a long-wavelength scalar field on a binary black hole spacetime. Owing to the inherent separation of scales, this problem is analytically tractable and a perturbative solution is obtained in Sec.~\ref{sec:pert}. The result is then discussed in two stages. In Sec.~\ref{sec:bound}, we track the evolution of a scalar cloud and calculate the rate at which different bound states grow or decay as a result of orbital superradiance. Then, in Sec.~\ref{sec:cont}, we show that the periodic forcing exerted by the binary  inevitably converts a fraction of these bound states into outgoing scalar radiation. Also in this section, we reanalyze the effect of mode mixing on the amplification of scalar waves. Finally, a brief summary of our key results is presented in Sec.~\ref{sec:discussion}. Supplementing this main line of discussion are Appendices~\mbox{\ref{sec:app_radial}--\ref{sec:app_mass_ratio}}, which contain additional technical details and some of the lengthier derivations.

\section{Effective field theory}
\label{sec:eft}

Our goal is to study the evolution of a real Klein--Gordon field $\phi(x)$ around a binary black hole. In realistic scenarios, the energy density in this field is always expected to be sufficiently dilute that its backreaction  may be neglected as a first approximation. Even in this test-field limit, however, obtaining a solution is prohibitively difficult for current analytic methods due to the complexity of the spacetime; hence, further assumptions are necessary to render the problem tractable.

In this paper, we restrict our attention to scalar-field configurations whose characteristic length scale~$\lambda$ is much greater than the binary's orbital separation~$a$. Because the individual black holes cannot be resolved by such a long-wavelength field, the binary as a whole behaves like an effective point particle that couples to the scalar via a set of multipole moments. This coarse-grained description of the system is essentially an extension of the multipolar post-Minkowskian formalism \cite{Blanchet:1985sp} with the addition of a~scalar~field.

The interactions between this effective point particle and the fields living in the bulk are encoded in the action \cite{Wong:2019kru}
\begin{equation}
	S_\text{pp} =
	-\int\dx\tau\, M
	+
	\sum_{\ell=0}^\infty \int\dx\tau
	\,O^L(\tau) \nabla_L \phi(\tau)
	+
	\cdots,
\label{eq:eft_S}
\end{equation}
where ${ M=M_1+M_2 }$ is the total mass of the binary. The scalar field ${\phi(\tau) \equiv \phi( z(\tau) )}$ is to be evaluated at the position of the binary's barycenter, which travels along the worldline~$z^\mu(\tau)$ with proper time given by ${ \dx\tau = \sqrt{- g_{\mu\nu} \dx x^\mu \dx x^\nu} }$. In the second term,%
\footnote{We use conventional multi-index notation:~A tensor with $\ell$ spatial indices is written as
${ O^L \equiv O^{i_1 \cdots i_\ell} }$,
while $\ell$ factors of a vector are written as
${ \bmf z^L \equiv \bmf z^{i_1} \cdots \bmf z^{i_\ell} }$,
and similarly
$\nabla_L \equiv \nabla_{i_1} \cdots \nabla_{i_\ell}$. Angled brackets around indices denote the STF projection of a tensor. In \eqref{eq:eft_S}, the indices ${ i \in \{1,2,3\} }$ label the three spatial directions that are mutually orthogonal to one another and to the tangent of the worldline $z^\mu(\tau)$. In the nonrelativistic limit, these reduce to the usual three spatial directions in Minkowski space.}
the composite operators $O^L(\tau)$ are symmetric and trace-free (STF) tensors localized on the worldline that capture how this point particle's internal degrees of freedom interact with the long-wavelength scalar \cite{Goldberger:2005cd, Endlich:2016jgc, Wong:2019yoc}. The ellipsis in \eqref{eq:eft_S} alludes to the presence of analogous composite operators that couple to the gravitational field \cite{Goldberger:2009qd}, although these will play no role in our discussion.

\paragraph{Induced multipole moments}

Generically, each of the composite operators in \eqref{eq:eft_S} can be decomposed into two pieces~\cite{Porto:2016pyg}. Writing ${ O^L = O^L_{(S)} + O^L_{(R)} }$, the first term is a \emph{permanent} multipole that is present if the internal degrees of freedom can directly source the scalar field. Such terms would be present, for instance, were the action in \eqref{eq:eft_S} to describe a binary neutron star system in a scalar--tensor theory of gravity \cite{Damour:1992we, Damour:1998jk, Alsing:2011er, Mirshekari:2013vb, Lang:2013fna, Sennett:2016klh, Kuntz:2019zef}. A binary black hole in general relativity, however, is not generic because stringent no-hair theorems \cite{Israel:1967wq, Bekenstein:1971hc, Hawking:1972qk, 1971ApJ...166L..35T, Adler:1978dp, Zannias:1994jf, Bekenstein:1995un, Saa:1996aw, Sotiriou:2011dz, Chrusciel:2012jk} stipulate that Kerr black holes cannot support stationary scalar-field configurations. Put another way, Kerr black holes cannot possess permanent scalar charges. Accordingly, ${O^L_{(S)} = 0}$ $\forall\,\ell$ in our case.~What remains is $O^L_{(R)}$, which represents the binary's \emph{dynamical} multipole moments that are induced in response to external perturbations. From now on, we drop the subscript~$(R)$ to declutter our notation.

As $\phi(x)$ is assumed to behave like a test field, the solution for these induced multipoles is given by linear response theory. With retarded boundary conditions imposed \cite{Endlich:2016jgc, Goldberger:2005cd, Porto:2016pyg},
\begin{equation}
	O^L(\tau) = \int \dx\tau'
	i \theta(\tau-\tau') \sum_{\ell'=0}^\infty
	\big\langle \big[ O^L(\tau), O^{L'}(\tau') \big] \big\rangle
	\nabla_{L'}\phi(\tau').
\label{eq:eft_O_response}
\end{equation}
Written in this quantum-mechanical language, the expectation value $\avg{\cdots}$ above requires the notion of some Hilbert space for the internal degrees of freedom. When specified correctly, this formalism can be used to systematically incorporate quantum effects like Hawking radiation~\cite{Goldberger:2019sya}. That being said, in this work we will only be interested in purely classical observables. It is then unnecessary to specify the density matrix with which this expectation value is taken, as the commutator is simply a $c$-number when Planck-suppressed terms are neglected~\cite{Goldberger:2019sya}.

By matching this post-Minkowskian formulation of the binary to a post-Newtonian description valid in its near zone, Ref.~\cite{Wong:2019kru} showed that the classical solution to \eqref{eq:eft_O_response} is
\begin{equation}
	O^L(t) = - \sum_{N=1}^2 \sum_{\ell'=0}^\infty \frac{A_N}{\ell!\ell'!}
	\bmf z_N^\avg{L}(t)
	\frac{\dx}{\dx t}\Big[ \bmf z_N^{\avg{L'}}(t) \partial_{L'} \phi(t,\bmf 0) \Big]
\label{eq:eft_O}
\end{equation}
in the nonrelativistic, low-frequency limit. In writing this solution, we have chosen coordinates such that the binary's barycenter is at rest at the origin. The motion of the $N$th black hole, whose horizon has area $A_N$, is then given by the vector~$\bmf z_N(t)$.

\paragraph{Expansion parameters}

At this stage, it is worth enumerating the four different expansion parameters that appear in this EFT. It will be convenient to assume that the two black holes have comparable masses for the purposes of power counting, although this formalism remains valid for arbitrary mass ratios as long as the binary's constituents are widely separated.

The first two expansion parameters come from the post-Newtonian description of the binary in its near zone: the solution in \eqref{eq:eft_O} is only the leading term in a series organized as an expansion in the binary's orbital velocity ${v \sim \sqrt{GM/a}}$ and the ratio of timescales $GM\w$. To be specific, the latter is the ratio of the black holes' light-crossing times to the characteristic time scale $\w^{-1}$ of the scalar.  The fact that \eqref{eq:eft_O} depends only on the areas of the black holes but not on their spins is a consequence of working to leading order in these parameters \cite{Wong:2019kru}. The induced multipoles are generated as a result of absorption of the scalar across the horizons \cite{Wong:2019yoc}, and a black hole's absorption cross section in the low-frequency limit is $s$-wave dominated and equal to its area~\cite{FHM}.

The approximation of the entire binary as an effective point particle introduces an additional two expansion parameters: the post-Minkowskian parameter $GM/\lambda$ and the ratio of length scales~$a/\lambda$. The first of these characterizes the nonlinearity of a given term in the solution due to self-interactions of the gravitational field. In this work, we work to first order in $GM/\lambda$ and will treat it nonperturbatively in order to capture the bound states of $\phi(x)$, but otherwise we neglect all higher-order corrections in $GM/\lambda$.

As for the fourth expansion parameter, we will---rather unusually from the point of view of an EFT---treat terms with different powers of $a/\lambda$ on equal footing, despite higher-order terms being parametrically suppressed. Doing so will allow us to keep track of the mixing between different angular-momentum modes, which leads to interesting consequences.

These four parameters control different aspects of the perturbative expansion, but enforcing the two conditions ${v \ll 1}$ and ${a/\lambda \ll 1}$ is often sufficient to ensure that we are inside the EFT's regime of validity. For a scalar cloud that is gravitationally bound to a binary, its characteristic frequency is set by the scalar field's mass, ${\w \simeq \mu}$, up to some nonrelativistic binding energy $\sim\! GM\mu/\lambda_\text{dB}$. In the denominator is the  de~Broglie wavelength ${\lambda_\text{dB} \sim (GM\mu^2)^{-1}}$, which determines the characteristic length scale of the cloud. This second expression may be used to recast the condition ${a/\lambda\ll 1}$ into an upper bound for the scalar's mass, namely,
\begin{equation}
	\frac{\mu}{\Omega} \ll \frac{1}{v^2},
\label{eq:eft_cutoff_mu}
\SetDisplaySkip[0][0.8]
\end{equation}
which follows after using ${v^2 \sim GM/a}$ and ${v^3 \sim GM\Omega}$. As a rough guide, \eqref{eq:eft_cutoff_mu} says that a scalar field should have a mass $\mu \ll 10^{-11}~\text{eV}\, (v/0.1)(M_\odot/M)$ if it is to engulf a binary of total mass $M$ with orbital velocity $v$.

An additional upper bound must be established for freely propagating scalar waves that impinge on the binary with frequency $\w$ and momentum $k \sim 1/\lambda$. For high-momentum modes, we choose to replace the necessary condition ${a/\lambda \ll 1}$ with the sufficient condition ${a\w \ll 1}$ for simplicity. The latter equivalently reads
\begin{equation}
	\frac{\w}{\Omega} \ll \frac{1}{v}.
\label{eq:eft_cutoff_w}
\SetDisplaySkip[0][0.8]
\end{equation}
For low-momentum modes, the quantity $a/\lambda$ can be arbitrarily small, so now the ultraviolet (UV) cutoff for this EFT is set by another expansion parameter, namely, $GM\w$. Since ${\w \sim \mu}$ in this limit, the condition ${GM\w \ll 1}$ is equivalent to ${\mu/\Omega \ll v^{-3}}$. This upper bound is weaker than that of \eqref{eq:eft_cutoff_mu}, and thus we are guaranteed to remain in the EFT's regime of validity when the UV cutoffs in \eqref{eq:eft_cutoff_mu} and \eqref{eq:eft_cutoff_w} are both respected.

\paragraph{Equation of motion}
Extremizing the total effective action with respect to $\phi(x)$, we obtain 
\begin{equation}
	\left( \Box - \mu^2 + \frac{2GM \mu^2}{r} \right) \phi(x)
	=
	- \sum_{\ell=0}^\infty (-1)^\ell O^L(t) \partial_L \delta^{(3)}(\bmf x),
\label{eq:eft_eom}
\end{equation}
where $\Box$ denotes the wave operator on flat space and we have also included the leading contribution from the binary's gravitational potential on the lhs. On the rhs, the induced multipoles $O^L(t)$ are given by \eqref{eq:eft_O}; hence, this is a linear, homogeneous differential equation for $\phi(x)$.

It is worth remarking that the delta function on the rhs of \eqref{eq:eft_eom} inevitably leads to singularities. Likewise, singularities also arise from the operators $O^L(t)$, which are functions of the scalar field and its derivatives evaluated at the origin. These UV divergences, which in this case originate from the point-particle approximation of the binary, are commonplace in EFTs and can be dealt with in the usual way by using a convenient regulator in conjunction with a renormalization scheme \cite{Goldberger:2004jt, Burgess:2016lal}. This procedure turns out to be unnecessary, however, up to first order in perturbation theory.

\section{Perturbative solution\protect\\via Green's function}
\label{sec:pert}

We obtain an approximate solution to \eqref{eq:eft_eom} by treating the interaction terms on the rhs as small perturbations. Denoting the differential operator on the lhs by $D_x$, this entails looking for a solution of the form ${\phi = \phi^{(0)} + \phi^{(1)} + \cdots\,}$, where the zeroth-order piece is an exact solution to the noninteracting theory, ${D_x \phi^{(0)}(x) = 0}$, while the first-order correction is
\begin{align}
	\phi^{(1)}(x)
	&=
	\int\dx^4x' G(x,x')\sum_{\ell'=0}^\infty (-1)^{\ell'}
	O^{(0)}_{L'}(t') \partial_{L'}^{\vphantom{(0)}} \delta^{(3)}(\bmf x')
	\nonumber\\
	&\quad+
	\phi^{(1)}_\text{cf}(x).
\label{eq:pert_phi_1}
\end{align}
The first term is the particular integral sourced%
\footnote{More accurately, the terms on the rhs of \eqref{eq:eft_eom} should be viewed as sink terms, since the induced multipoles arise from absorption.}
by the induced multipoles $O^{(0)}_L(t)$, where the superscript~$(0)$ indicates that the expression in \eqref{eq:eft_O} is to be evaluated using the zeroth-order solution~$\phi^{(0)}$. Meanwhile, the second term in \eqref{eq:pert_phi_1} is the complementary function, ${D_x \phi^{(1)}_\text{cf}(x)=0}$, whose inclusion may be necessary to ensure that the overall solution satisfies our choice of boundary conditions.

In this section, we discuss the three ingredients that make up the particular integral. We begin by writing down the general solution $\phi^{(0)}$ to the noninteracting theory, which is then fed into \eqref{eq:eft_O} to obtain explicit expressions for $O^{(0)}_L(t)$. Finally, taking their convolution with the retarded Green's function $G(x,x')$ produces the end result.

\subsection{The noninteracting theory}
\label{sec:pert_0}

To establish some nomenclature and introduce the basis for our perturbative approach, we begin in this subsection with a brief review of the Coulomb functions \cite{HUMBLET1984,BetheSalpeter}.

As the noninteracting theory is time-translation invariant and spherically symmetric, we may look for solutions of the form
${ \phi(x) \propto R(r) Y_{\ell m}(\hat{\bmf x}) e^{-i\w t} }$,
where $Y_{\ell m}(\hat{\bmf x})$ are the usual spherical harmonics. The radial part $R(r)$ must then satisfy the differential equation
\begin{equation}
	\left( \frac{\dx^2}{\dx r^2} + k^2 + \frac{2GM\mu^2}{r} - \frac{\ell(\ell+1)}{r^2}\right)
	r R(r) = 0
\label{eq:pert_eom_R}
\end{equation}
with ${ k^2 \equiv \w^2 - \mu^2 }$. The resulting set of solutions can be divided into three categories depending on the value of this quantity. In what follows, we define ${ k\equiv k(\w) }$ as the appropriate root of $k^2$, namely
\begin{equation}
\SetDisplaySkip[0]
\AddDisplaySkip[0.3]
	k(\w) \coloneq
	\begin{cases}
		\text{sgn}(\w) \sqrt{\w^2 - \mu^2} & (k^2 \geq 0),
		\\[0.5em]
		i\sqrt{\mu^2 - \w^2} & (k^2 < 0).
	\end{cases}
\label{eq:pert_k}
\end{equation}

\paragraph{Radiation modes}
Let us begin with the case ${k^2 \geq 0}$. Defining ${\zeta \coloneq - GM\mu^2/k}$, two linearly independent solutions to \eqref{eq:pert_eom_R} are
\begin{subequations}
\label{eq:pert_R}
\begin{equation}
\SetDisplaySkip[0]
	R^\pm_\ell(k,r)
	\coloneq
	\frac{H_\ell^\pm(\zeta,kr)}{\pm ikr},
\end{equation}
where $H_\ell^\pm$ are Coulomb functions. (We follow the conventions in Ref.~\cite{DLMF}.) From their asymptotic forms at large $r$, given in \eqref{eq:app_H_asymptotic}, we can deduce that these solutions correspond to ingoing~($-$) and outgoing~($+$) spherical waves.

For later purposes, it will also be useful to define a particular linear combination of these \emph{radiation modes}. Let
\begin{equation}
\label{eq:pert_R0}
	R_\ell(k,r)
	\coloneq \frac{1}{2} [ R^+_\ell(k,r) + R^-_\ell(k,r) ]
	\equiv \frac{F_\ell(\zeta,kr)}{kr},
\end{equation}
\end{subequations}
where $F_\ell$ is another Coulomb function. This solution describes a superposition of ingoing and outgoing waves in equal measure and is regular at the origin as a result. Additionally, let us define the mode functions
\begin{subequations}
\label{eq:pert_phi_mf}
\begin{align}
	\phi^\pm_{k\ell m}(\bmf x)
	&\coloneq
	R^\pm_\ell(k,r) Y_{\ell m}(\hat{\bmf x}),
	\\[0.1em]
	\phi_{k\ell m}(\bmf x)
	&\coloneq
	R_\ell(k,r) Y_{\ell m}(\hat{\bmf x}),
\end{align}
\end{subequations}
which describe the three-dimensional spatial profile of these scalar waves.

\paragraph{Yukawa modes}
Solutions to \eqref{eq:pert_eom_R} for the case ${k^2<0}$ may now be obtained by analytic continuation. Represented in terms of Whittaker functions~\cite{DLMF}, they read
\begin{subequations}
\label{eq:app_R_analytic_cont}
\begin{align}
	\rule{0pt}{18pt}
	R^\pm_\ell(k,r)
	&=
	(\mp i)^{\ell+1} e^{\pi\zeta/2 \pm i\sigma_\ell(\zeta)}
	\frac{ W_{\mp i\zeta,\ell+1/2}(\mp2ikr) }{kr},
	\\
	R_\ell(k,r)
	&=
	\frac{ C_\ell(\zeta) }{ (-2i)^{\ell+1} }
	\frac{ M_{-i\zeta,\ell+1/2}(-2ikr) }{kr},
\end{align}
\end{subequations}
where the Coulomb phase shift $\sigma_\ell(\zeta)$ and the Gamow factor $C_\ell(\zeta)$ are given by~\cite{HUMBLET1984,DLMF}
\begin{subequations}
\begin{gather}
	\sigma_\ell(\zeta)
	\coloneq
	\frac{1}{2i} [\log\Gamma(\ell+1+i\zeta) - \log\Gamma(\ell+1-i\zeta)],
	\label{eq:pert_Coulomb_phase_shift}
	\allowdisplaybreaks\\
	C_\ell(\zeta)
	\coloneq
	\frac{\Gamma(\ell+1-i\zeta)}{\Gamma(2\ell+2)} 2^\ell
	e^{i\sigma_\ell(\zeta) - \pi\zeta/2}.
\end{gather}
\end{subequations}

For imaginary $k$ defined according to \eqref{eq:pert_k}, the $R^-_\ell(k,r)$ solution is seen to grow exponentially with~$r$ and is therefore unphysical. In contrast, $R^+_\ell(k,r)$ describes a nonpropagating field profile with characteristic size ${\lambda \sim 1/|k|}$. Depending on the value of ${k \in i\mathbb R_{>0}}$, this solution can either be singular or regular at ${r=0}$.

The set of singular solutions includes the ${\w=\ell=0}$ mode, which has the asymptotic~form
\begin{equation}
	R^+_0(i\mu,r) \sim \frac{e^{-\mu r}}{r}
	e^{GM\mu \log(2\mu r)}[ 1 + \mathcal O(r^{-1})]
\end{equation}
at large $r$ up to some constant prefactor. One easily recognizes this as the Yukawa potential sourced by a point charge at the origin, albeit with corrections coming from the gravitational potential of the point mass $M$. In general, we will refer to this set of singular solutions as the \emph{Yukawa modes}. Because these modes correspond to having pure imaginary $k$ in the continuous domain ${i\mathbb R_{>0} }$ modulo a discrete set of points to be discussed below, the term \emph{continuum states} will be used to refer to the combined set of radiation modes and Yukawa modes.

\paragraph{Bound states}
The solution $R^+_\ell(k,r)$ is regular at the origin when $k$ takes special values such that ${-i\zeta = n}$ is an integer and $n \geq \ell+1$. These regular solutions have corresponding frequencies $\w = \pm E_n$ given by
\begin{equation}
	E_n =
	\mu \left( 1 - \frac{(GM\mu)^2}{n^2}\right)^{1/2}
	\!\simeq
	\mu - \frac{\mu (GM\mu)^2}{2n^2},
\end{equation}
which is reminiscent of the hydrogen bound-state spectrum. Accordingly, this set of regular solutions will be called the \emph{bound states}. To make the connection to the hydrogen atom even more explicit, let us define $R_{n\ell}(r)$ to be a rescaled version of the solution $R^+_\ell(k,r)$ when evaluated at ${\zeta=in}$.%
\footnote{This rescaling is also necessary on mathematical grounds, as the original solution $R^+_\ell(k,r)$ vanishes when ${\zeta=in}$. This behavior can be traced back to the Coulomb phase shift, as the first gamma function in \eqref{eq:pert_Coulomb_phase_shift} is being evaluated at one of its poles.} 
In terms of the Whittaker functions, it reads
\begin{align}
	R_{n\ell}(r)
	&\coloneq
	\frac{ (-1)^{n-\ell-1} (-2ik)^{3/2}}%
	{\sqrt{2n (n+\ell)! (n-\ell-1)!}}
	\frac{ W_{-i\zeta,\ell+1/2}(-2ikr) }{-2ikr}
	\bigg|_{\zeta=in}
	\nonumber\\[0.5em]
	&\equiv
	\sqrt{\frac{(n+\ell)! (-2ik)^3}{2n (n-\ell-1)!}}
	\frac{ M_{-i\zeta,\ell+1/2}(-2ikr) }{(2\ell+1)!(-2ikr)}
	\bigg|_{\zeta=in}.
\label{eq:app_Rnl_Whittaker}
\end{align}
The second line follows from Eq.~(13.14.32) of Ref.~\cite{DLMF}, which shows that $R^+_\ell$ and $R_\ell$ become proportional to one another when ${\zeta \to in}$. Note also that ${-ik = GM\mu^2/n}$ in this limit, and thus $R_{n\ell}(r)$ is a real-valued function. With this definition, the mode functions
\begin{equation}
	\psi_{n\ell m}(\bmf x) \coloneq R_{n\ell}(r) Y_{\ell m}(\hat {\bmf x})
\end{equation}
are exactly the (orthonormal) hydrogen wave functions, albeit with $GM\mu$ in place of the fine-structure constant.

\paragraph{General solution}
Because we are treating absorption of the scalar field by the black holes perturbatively via interaction terms, the origin is devoid of sinks or sources in the noninteracting theory. Consequently, the zeroth-order solution $\phi^{(0)}$ must be regular at ${r=0}$. This boundary condition precludes the existence of Yukawa modes at this order and, moreover, a net flux of radiation into or out of the origin is also prohibited. The general solution is thus given by the linear combination
\begin{align}
	\phi^{(0)}(x)
	&=
	\sum_{\ell,m}\int\frac{\dx\w}{2\pi} 2\mathcal I^>_{\w\ell m} 
	\phi_{k\ell m}(\bmf x) e^{-i\w t}
	\nonumber\\
	&\quad+
	\frac{1}{\sqrt{2\mu}} \sum_{n,\ell,m} c^{(0)}_{n\ell m}
	\psi_{n\ell m}(\bmf x) e^{-i\w t}
	+
	\text{c.c.}
\label{eq:pert_phi_0}
\end{align}
The first term is the sum over a superposition of ingoing and outgoing waves [the factor of 2 follows from \eqref{eq:pert_R0} and \eqref{eq:pert_phi_mf}], where a given mode $(\w,\ell,m)$ has an ingoing amplitude specified by the coefficient $\mathcal I^>_{\w\ell m}$. The ``$>$'' symbol is used to emphasize that this function can be chosen without loss of generality to have support only in the domain ${\w \geq 0}$, corresponding to positive-frequency modes. The negative-frequency modes are then automatically taken into account by the complex conjugate ($\text{c.c.}$) terms. Further note that $\mathcal I^>_{\w\ell m}$ must vanish for ${\w \in (-\mu,\mu)}$ as per the boundary conditions described above. Meanwhile, the second term in \eqref{eq:pert_phi_0} is the sum over bound states, with a conventional prefactor of $1/\sqrt{2\mu}$ included to render the coefficients $c^{(0)}_{n\ell m}$ dimensionless.

\subsection{Induced multipoles for circular orbits}
\label{sec:pert_Olm}

The zeroth-order solution in \eqref{eq:pert_phi_0} may now be fed into \eqref{eq:eft_O} to obtain the binary's induced multipoles. To that end, we begin by introducing some compact notation: let ${ \bs \equiv (n,\ell,m) }$ collectively refer to the three integers that specify a bound state and, likewise, let ${ \cs \equiv (\w,\ell,m) }$ refer to the parameters for a given continuum state. We then write
\begin{equation}
	\sum_\bs \equiv \sum_{n,\ell,m}
	\quad\text{and}\quad
	\sum_\cs \equiv \sum_{\ell,m} \int\frac{\dx\w}{2\pi}
\label{eq:pert_sum_def}
\end{equation}
to denote summing over the bound and continuum states, respectively. In this new shorthand, the zeroth-order solution reads
\begin{align}
	\phi^{(0)}(x)
	&=
	\sum_\cs
	2\mathcal I^>_\cs \phi_{k\ell m}(\bmf x)
	e^{-i\w t}
	+
	\frac{1}{\sqrt{2\mu}}\sum_\bs
	c_\bs^{(0)} \psi_\bs(\bmf x) e^{-i E_n t}
	\nonumber\\&\quad
	+
	\text{c.c.}
\label{eq:pert_phi_0_short}
\end{align}

Rather than substitute this directly into \eqref{eq:eft_O} to produce a set of tensorial objects, it is more convenient to work with the components of $O^L(t)$ obtained via projection onto a basis of STF tensors. For each $\ell$, let $ \Y^{\ell m}_L$ denote the basis vectors that generate the spherical harmonics [i.e., ${Y_{\ell m}(\hat{\bmf x}) = \Y^{\ell m}_L \hat{\bmf x}^L}$] and satisfy the orthogonality relation \cite{Thorne:1980ru}
\begin{gather}
	(\Y^{\ell m}_L)^* \Y^{\ell m'}_L = \frac{(2\ell+1)!!}{4\pi\ell!} \delta^{mm'}.
\label{eq:pert_Y_orthogonal}
\end{gather}
The $2\ell+1$ independent degrees of freedom of $O^L(t)$ can then be obtained via the projection
\begin{equation}
	O_{\ell m}(t) = -\frac{1}{4\pi i} \frac{4\pi\ell!}{(2\ell+1)!!}(\Y^{\ell m}_L)^* O^L(t),
\end{equation}
where the prefactor of $-4\pi i$ is included purely for convenience. To reconstruct $O^L(t)$, one simply inverts this relation to find
\begin{equation}
	O^L(t) = - 4\pi i \sum_{m=-\ell}^\ell \Y^{\ell m}_L O_{\ell m}(t).
\label{eq:pert_OL}
\end{equation}

At the moment, the formula in \eqref{eq:eft_O} for these induced multipoles does not make any assumptions about the black holes' trajectories, apart from requiring that ${|\bmf z_N(t)| \ll \lambda}$. For simplicity, in this paper we will restrict our attention to circular orbits with frequency $\Omega$ oriented such that its angular momentum points along the positive $z$~axis. For this configuration, Ref.~\cite{Wong:2019kru} showed that the components $O_{\ell m}(t)$ are given by
\begin{align}
	O_{\ell m}(t)
	&= \sum_{\ell',m'}
	\frac{ 4\pi Y^*_{\ell m}(\bmf d) Y_{\ell'm'}(\bmf d) B_{\ell\ell'} }{(2\ell+1)!! (2\ell'+1)!!}
	e^{-i(m-m')\Omega t}
	\nonumber\\&\quad\times
	(m'\Omega -i\partial_t)(\Y^{\ell'm'}_{L'})^*
	\partial_{L'}\phi(t,\bmf 0),
\label{eq:pert_Olm_circular}
\end{align}
where $\bmf d$ is the unit vector parallel to $\bmf z_1(0)$ and
\begin{equation}
	B_{\ell\ell'} \coloneq \sum_N A_N r_N^{\ell+\ell'}
\label{eq:pert_Bll}
\end{equation}
characterizes the interaction strength between the black holes and the scalar, with ${r_1 = a M_2/M}$ and ${r_2 =-a M_1/M}$ denoting the displacements of the black holes from their barycenter.

The induced multipoles $O^L(t)$ are necessarily real by construction; hence, the definition in \eqref{eq:pert_OL} can be used to show that its components must satisfy the constraint
\begin{equation}
	O_{\ell m}^*(t) = -(-1)^{m} O_{\ell,-m}(t),
\label{eq:pert_Olm_constraint}
\end{equation}
which follows from the identity for the complex conjugate of a spherical harmonic; cf.~\eqref{eq:app_cc_Y}. This motivates writing
\begin{equation}
	O_{\ell m}(t) = O_{\ell m}^>(t) - (-1)^m O_{\ell,-m}^{>*}(t)
\label{eq:pert_Olm_split}
\end{equation}
such that \eqref{eq:pert_Olm_constraint} is automatically satisfied for any function $O_{\ell m}^>(t)$. It is then easy to show that for a real scalar-field solution of the form ${\phi(x) = \phi^>(x) + \text{c.c.}}$, one obtains $O_{\ell m}^>(t)$ by evaluating the rhs of \eqref{eq:eft_O} using only $\phi^>(x)$ rather than $\phi(x)$. In other words, we find
\begin{widetext}
\begin{align}
	O_{\ell m}^{(0)>}(t)
	&=
	\sum_{\cs'}
	\frac{Y^*_{\ell m}(\bmf d) Y_{\ell'm'}(\bmf d) B_{\ell\ell'}}{(2\ell+1)!!}
	\mathfrak R_{\ell'}(k') 
	(m'\Omega - \w') 2\mathcal I^>_{\cs'}
	e^{-i[\w' + (m-m')\Omega]t}
	\nonumber\\&\quad
	+
	\frac{1}{\sqrt{2\mu}} \sum_{\bs'}
	\frac{Y^*_{\ell m}(\bmf d) Y_{\ell'm'}(\bmf d) B_{\ell\ell'}}{(2\ell+1)!!}
	\mathfrak R_{n'\ell'}
	(m'\Omega - E_{n'}) c_{\bs'}^{(0)}
	e^{-i [E_{n'} + (m-m')\Omega]t}
\label{eq:pert_Olm_explicit}
\end{align}
\end{widetext}
after substituting only the positive-frequency part of \eqref{eq:pert_phi_0_short} into \eqref{eq:pert_Olm_circular}. In obtaining this result, we have made use of the identities in \eqref{eq:app_derivatives_mode_functions} and have defined
\begin{align}
	\mathfrak R_\ell(k)
	&\coloneq
	\lim_{r\to 0}\frac{1}{\ell!}
	\frac{\dx^\ell}{\dx r^\ell} R_\ell(k,r),
	\\
	\mathfrak R_{n\ell}
	&\coloneq
	\lim_{r\to 0}\frac{1}{\ell!}
	\frac{\dx^\ell}{\dx r^\ell} R_{n\ell}(r),
\end{align}
explicit expressions for which are provided in \eqref{eq:app_derivative_R}. The result in \eqref{eq:pert_Olm_explicit} can now be used to determine the first-order correction $\phi^{(1)}$ via the method of Green's functions.

\subsection{Integration contours}
The Green's function in \eqref{eq:pert_phi_1} is defined by the equation
${ D_x G(x,x') = - \delta^{(4)}(x-x') }$ and may be written as the inverse Fourier transform
\begin{equation}
	G(x,x') =
	\sum_{\ell,m}\int\frac{\dx\w}{2\pi}\, 
	G_{\w\ell}(r,r') Y_{\ell m}(\hat{\bmf x}) Y_{\ell m}^*(\hat{\bmf x}') e^{-i\w(t-t')},
\end{equation}
where the radial part is given by \cite{Hostler,Mapleton}
\begin{align}
	G_{\w\ell}(r,r') &=
	(-2ik)\frac{\Gamma(\ell+1+i\zeta)}{(2\ell+1)!}
	\frac{ W_{-i\zeta,\ell+1/2}(-2ikr_>) }{-2ik r_>}
	\nonumber\\&\quad\times
	\frac{ M_{-i\zeta,\ell+1/2}(-2ikr_<) }{-2ik r_<}
\end{align}
with $r_> \coloneq \max(r,r')$ and $r_< \coloneq \min(r,r')$.

To evaluate the particular integral in \eqref{eq:pert_phi_1}, we first note that the delta function imposes the restrictions ${r_< = r'}$ and ${r_> = r}$, which make performing the integral over $\bmf x'$ relatively straightforward.  First integrating by parts to move the derivatives $\partial_{L'}$ onto $G(x,x')$, we obtain
\begin{align}
	\phi^{(1)}(x)
	&\supset
	\sum_{\ell,m}\int\frac{\dx\w}{2\pi}
	\sum_{\ell'=0}^\infty \int\dx^4x'
	O^{(0)}_{L'}(t') e^{-i\w(t-t')}
	\nonumber\\
	&\quad\times
	(-2ik)\frac{\Gamma(\ell+1+i\zeta)}{(2\ell+1)!}
	\frac{ W_{-i\zeta,\ell+1/2}(-2ikr) }{-2ik r}
	Y_{\ell m}(\hat{\bmf x})
	\nonumber\\
	&\quad\times
	\delta^{(3)}(\bmf x') \partial_{L'}
	\left(
		\frac{ M_{-i\zeta,\ell+1/2}(-2ikr') }{-2ik r'}
		Y_{\ell m}^*(\hat{\bmf x}')
	\right).
\end{align}
Now using \eqref{eq:pert_Y_orthogonal}, \eqref{eq:pert_OL}, and \eqref{eq:app_derivative_Whittaker_M}, this can be shown to simplify to
\bgroup\predisplaypenalty=0
\begin{align}
	\phi^{(1)}(x)
	&\supset
	-i\sum_{\ell,m}\int\frac{\dx\w}{2\pi}
	\frac{ \Gamma(\ell+1+i\zeta) }{(2\ell)!!}
	\int\dx t' O_{\ell m}^{(0)}(t') e^{-i\w(t-t')}
	\nonumber\\&\quad\times
	(-2ik)^{\ell+1} \frac{ W_{-i\zeta,\ell+1/2}(-2ikr) }{-2ikr}
	Y_{\ell m}(\hat{\bmf x})
\label{eq:pert_PI_bound_exp}
\end{align}
\egroup
after also using the identity ${n! \equiv n!! (n-1)!!}$ for the double factorial. Given that ${(2\ell)!! = 2^\ell \ell!}$, an equivalent expression for the particular integral is
\begin{align}
	\phi^{(1)}(x)
	&\supset
	\sum_{\ell,m}\int\frac{\dx\w}{2\pi}
	\frac{ \Gamma(\ell+1+i\zeta) }{\Gamma(\ell+1)}
	\int\dx t' O_{\ell m}^{(0)}(t') e^{-i\w(t-t')}
	\nonumber\\&\quad\times
	(-ik)^{\ell+1}
	\frac{ W_{-i\zeta,\ell+1/2}(-2ikr) }{kr}
	Y_{\ell m}(\hat{\bmf x}).
\label{eq:pert_PI_cont_exp}
\end{align}
Both expressions will turn out to be useful in later sections.

\begin{figure}[b]
\centering\includegraphics{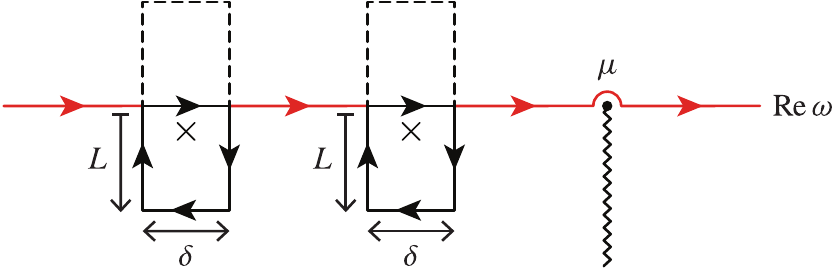}
\caption{Illustration (not to scale) showing the different contributions to the $\w$ integral. The principal value (red line) gives rise to the continuum states, while the bound states come from closed contours (black lines) that encircle each of the poles of the Green's function. The poles (cross marks) are shifted off the real axis to enforce retarded boundary conditions and thus contribute to the solution only when the black contours are closed in the lower half of the complex plane. The limit $\delta \to 0$ and $L \to \infty$ should be taken at the end. Note also the presence of branch points at $\w=\pm\mu$ and at infinity.}
\label{fig:contour}
\end{figure}

It remains to perform the integrals over $t'$ and $\w$. Care must be exercised with the latter because the Green's function contains poles at ${\w = \pm E_n}$ and branch points at ${\w = \pm \mu}$ and at infinity \cite{Hostler}. To proceed, we split the integral over $\w$ into two parts: its principal value along the real line gives rise to the continuum states, while the bound states are obtained by integrating over closed contours encircling each of the poles; see also Fig.~\ref{fig:contour}.

The complete first-order solution is thus a sum of three parts:
\begin{equation}
	\phi^{(1)}_{\vphantom{b}}
	= 
	\phi^{(1)}_b + \phi^{(1)}_{c\vphantom{h}} + \phi^{(1)}_\text{cf}.
\end{equation}
The first term contains the bound states, which we study in Sec.~\ref{sec:bound}, while the second contains the continuum states, discussed in Sec.~\ref{sec:cont}.  Finally, recall from our earlier discussion that the third term is a complementary function that may be required to satisfy boundary conditions.

\section{Bound states}
\label{sec:bound}

The bound states surrounding a binary evolve in an intricate manner as a result of their interaction with the black holes. In perturbation theory, this evolution can be regarded as being sourced by the induced multipoles $O^L(t)$. The first-order result $\phi_b^{(1)}$ is obtained by performing the $\w$ integral in \eqref{eq:pert_PI_bound_exp} over closed contours that each encircle one of the poles ${\w = \pm E_n}$ of the Green's function.

When $t>t'$, these contours should be closed in the lower half of the complex plane; see Fig.~\ref{fig:contour}. In the limit ${\delta \to 0}$ and ${L \to \infty}$, the integrals along the vertical paths cancel each other, while the integral over the bottom horizontal path vanishes. The residue theorem can then be applied to show that 
\begin{align}
	\phi^{(1)}_b(x)
	&=
	\sum_{s=\pm 1}\sum_\bs (-i)^2
	\frac{ \text{Res}[\Gamma(\ell+1+i\zeta)]}{(2\ell)!!} (-2ik)^{\ell+1}
	\nonumber\\&\quad\times
	\int\dx t' \theta(t-t') O_{\ell m}^{(0)}(t') e^{i\w t'}
	\nonumber\\&\quad\times
	\frac{ W_{-i\zeta,\ell+1/2}(-2ikr) }{-2ikr} Y_{\ell m}(\hat{\bmf x}) e^{-i\w t}
	\bigg|_{\w = s E_n - i\epsilon},
\label{eq:bound_phi_1_raw_1}
\end{align}
where the poles have been shifted by an amount $-i\epsilon$ to enforce retarded boundary conditions and the sum over ${s=\pm 1}$ is used to account for both the positive- and negative-frequency solutions. Note also the presence of the step function $\theta(t-t')$, which follows from the fact that the contour should be closed in the upper half of the complex plane when ${t<t'}$.

To determine the residue of the gamma function at the pole ${\w=sE_n}$, we use the standard Laurent expansion
$\Gamma(-j + z) = (-1)^j / ( j! z) + \mathcal O(z^0)$
valid for any non-negative integer $j$ to obtain
\begin{equation}
	\text{Res}[\Gamma(\ell+1+i\zeta)]_{\w=sE_n}
	=
	-\frac{(-1)^{n-\ell-1}(GM\mu^2)^2}{(n-\ell-1)! n^3 s E_n}.
\end{equation}
Substituting this back into \eqref{eq:bound_phi_1_raw_1} and using \eqref{eq:app_Rnl_Whittaker} to rewrite the Whittaker function in terms of $R_{n\ell}(r)$, one finds
\begin{align}
	\phi^{(1)}_b(x)
	&=
	\sum_{s=\pm 1}\sum_\bs
	\frac{\mathfrak R_{n\ell}}{2 s E_n} (2\ell+1)!! \psi_\bs(\bmf x)
	\nonumber\\&\quad\times
	\int\dx t' \theta(t-t')
	O_{\ell m}^{(0)}(t') e^{-i(s E_n - i\epsilon)(t-t')},
\end{align}
which can be further simplified to
\begin{align}
	\phi^{(1)}_b(x)
	&=
	\sum_\bs
	\frac{\mathfrak R_{n\ell}}{2 E_n} (2\ell+1)!! \psi_\bs(\bmf x)
	\int^t\dx t'  O_{\ell m}^{(0)}(t') e^{-iE_n(t-t')}
	\nonumber\\&\quad
	+
	\text{c.c.}
\label{eq:bound_phi_1_raw_2}
\end{align}
The latter expression follows after making two observations. First, the ${s=-1}$ terms are exactly the complex conjugates of the ${s=+1}$ terms, which one can show by using the identities in \eqref{eq:pert_Olm_constraint} and \eqref{eq:app_cc_psi} together with the freedom to relabel ${m \to -m}$ as it is being summed over. Second, the lower bound of the $t'$ integral at $-\infty$ yields no contribution because of the $-i\epsilon$ term in the exponent, and thus we need only keep track of the result from the upper bound. Discarding this lower bound constitutes no loss in generality, as the freedom to specify initial conditions for the amplitudes of the bound states at some initial time, say ${t=0}$, is provided by our freedom to choose the complementary function $\phi^{(1)}_\text{cf}$.

The result in \eqref{eq:bound_phi_1_raw_2} is not yet in a useful form because $O^{(0)}_{\ell m}(t')$ contains both the positive- and negative-frequency parts of the zeroth-order solution, whose separate contributions we would like to make manifest. To do this, we use the decomposition in \eqref{eq:pert_Olm_split} to write
\begin{align}
	\phi^{(1)}_b(x)
	&=
	\sum_\bs \frac{\mathfrak R_{n\ell}}{2E_n} (2\ell+1)!! \,\psi_\bs(\bmf x)\!
	\int^t\dx t'
	\big[ O^{(0)>}_{\ell m}(t')
	\nonumber\\&\quad
	- (-1)^m O_{\ell,-m}^{(0)>*}(t') \big]  
	e^{-iE_n(t-t')}
	+\text{c.c.}
\end{align}
The freedom to swap the terms involving $O_{\ell,-m}^{(0)>*}(t)$ with their complex conjugates and to relabel ${m \to -m}$ gives us our final expression:
\begin{align}
	\phi^{(1)}_b(x)
	&=
	\sum_\bs \frac{\mathfrak R_{n\ell}}{2E_n} (2\ell+1)!! \,\psi_\bs(\bmf x)\!
	\int^t\dx t'
	\big[ O^{(0)>}_{\ell m}(t')
	\nonumber\\&\quad
	- O_{\ell m}^{(0)>}(t) e^{2iE_n (t-t')} \big]  
	e^{-iE_n(t-t')}
	+\text{c.c.}
\end{align}

It is now apparent that the full solution
${\phi = \phi^{(0)} + \phi^{(1)} + \cdots}$
for the bound states has the form
\begin{equation}
	 \phi_b(x)
 	=
 	\frac{1}{\sqrt{2\mu}} \sum_\bs
 	[ c_\bs(t) \psi_\bs(\bmf x) e^{-iE_n t} + \text{c.c.} ]
\label{eq:bound_phi_b_exact}
\end{equation}
with
${ c_\bs(t) \equiv c_\bs^{(0)} + c^{(1)}_\bs(t) + c^{(1)}_{\text{cf},\bs} + \cdots\, }$. The time evolution of the amplitude is given to first order by
\begin{align}
	c^{(1)}_\bs(t)
	&=
	\sqrt{2\mu}\frac{\mathfrak R_{n\ell}}{2E_n} (2\ell+1)!!
	\int^t\dx t' O^{(0)>}_{\ell m}(t')
	\nonumber\\&\quad\times
	\big( e^{i E_n t'} - e^{2i E_n t} e^{-i E_n t'} \big) ,
\label{eq:bound_c1}
\end{align}
while $c^{(1)}_{\text{cf},\bs}$ is a constant term coming from $\phi^{(1)}_\text{cf}$ that we tune in order to choose initial conditions.

Finally, substituting the expression for $O^{(0)>}_{\ell m}(t)$ in \eqref{eq:pert_Olm_explicit} into \eqref{eq:bound_c1} yields the explicit solution
\begin{widetext}
\begin{align}
	c^{(1)}_\bs(t)
	&=
		\sum_{\bs'} \frac{ V_{\bs\bs'} }{2E_n}
	(m'\Omega-E_{n'}) c_{\bs'}^{(0)}
	\int^t\dx t'
	\big( e^{i\Delta_{\bs\bs'}t'} - e^{2i E_n t} e^{i(\Delta_{\bs\bs'} - 2 E_n)t'} \big)
	\nonumber\\[0.2em]
	&\quad+
	\frac{1}{2\mu} \sum_{\cs'}
	\frac{ V_{\bs\cs'} }{ 2E_n } (m'\Omega-\w')
	2\mathcal I^>_{\cs'}
	\int^t\dx t'
	\big( e^{i\Delta_{\bs\cs'}t'} -	e^{2i E_n t} e^{i(\Delta_{\bs\cs'} - 2E_n)t'} \big),
\label{eq:bound_c1_exp}
\end{align}
\end{widetext}
which is written in terms of the energy differences
\begin{subequations}
\begin{align}
	\Delta_{\bs\bs'}
	&=
	E_n - E_{n'} - (m-m')\Omega,
	\label{eq:bound_Delta_bb}
	\\[0.2em]
	\Delta_{\bs\cs'}
	&=
	E_n - \w' - (m-m')\Omega
\end{align}
\end{subequations}
and the matrix elements
\begin{subequations}
\label{eq:bound_V}
\begin{align}
	V_{\bs\bs'}
	&=
	Y^*_{\ell m}(\bmf d)Y_{\ell'm'}(\bmf d) B_{\ell\ell'}
	\mathfrak R_{n\ell}\mathfrak R_{n'\ell'},
	\\
	V_{\bs\cs'}
	&=
	(2\mu)^{3/2} Y^*_{\ell m}(\bmf d)Y_{\ell'm'}(\bmf d) B_{\ell\ell'}
	\mathfrak R_{n\ell} \mathfrak R_{\ell'}(k').
\end{align}
\end{subequations}
Note the extra normalization factor of $1/(2\mu)$ in the second line of \eqref{eq:bound_c1_exp} has been included because the sum over continuum states $\sum_\cs$ is dimensionful; cf.~\eqref{eq:pert_sum_def}. Written in this way, $V_{\bs\bs'}$ and $V_{\bs\cs'}$ both have dimensions of energy.

\subsection{Growth rates}
We are now in a position to discuss the physical implications of this result. To begin with, suppose that ${c^{(0)}_\bs \neq 0}$ and consider its contribution to $c^{(1)}_\bs(t)$. Since ${\Delta_{\bs\bs} = 0}$, we have that
\begin{align}
	c^{(1)}_\bs(t)
	&\supset
	\frac{V_{\bs\bs}}{2E_n} (m\Omega-E_n) c^{(0)}_\bs
	\int^t \dx t' \big( 1- e^{2iE_n (t-t')} \big)
	\nonumber\\
	&=
	c^{(0)}_\bs \Gamma_\bs t + \text{const.}
\label{eq:bound_linear_growth}
\end{align}
Because the constant term may be removed by an appropriate choice of $c^{(1)}_{\text{cf},\bs}$, the physical effect of the diagonal element $V_{\bs\bs}$ is to cause the bound state $\bs$ to grow at the rate
\begin{equation}
	\Gamma_\bs \coloneq \frac{V_{\bs\bs}}{2E_n}(m\Omega-E_n).
\end{equation}
Thus, the bound states of the noninteracting theory turn into quasibound states once their interaction with the binary is taken into account. 

It is worth remarking that the linear growth in \eqref{eq:bound_linear_growth} is an approximation that is valid only at early times ${t \ll 1/\Gamma_\bs}$. Once $\Gamma_\bs t$ becomes of order unity, terms of the form ${\sim (\Gamma_\bs t)^p}$ that appear at higher orders in perturbation theory all become relevant. One might naturally expect that resumming these polynomials to all orders will lead to an exponentially growing solution $c_\bs(t) \propto \exp(\Gamma_\bs t)$ and, indeed, this turns out to be the case. The details of this resummation procedure, while interesting on theoretical grounds, have been relegated to Appendix~\ref{sec:app_resum} as they will not be relevant to this paper's main line of discussion. In what follows, it will suffice to work with the linear approximation in \eqref{eq:bound_linear_growth}.

Written out explicitly, the growth rate for the ${\bs\equiv (n,\ell,m)}$ mode reads
\begin{equation}
	\Gamma_\bs
	=
	\left|\frac{Y_{\ell m}(\bmf d)}{(2\ell+1)!}\right|^2
	\frac{(n+\ell)! (2GM\mu^2)^{2\ell+3}}{(n-\ell-1)! 4n^{2\ell+4} E_n}
	B_{\ell\ell} (m\Omega - E_n).
\label{eq:bound_growth_rate}
\end{equation}
It is instructive to first compare this result with the growth rate of a long-wavelength scalar cloud around a \emph{single} rotating black hole \cite{Detweiler:1980uk, Baumann:2019eav, Endlich:2016jgc}. Strikingly, after identifying the total mass $M$ of the binary with the mass of the single black hole and, likewise, identifying the binary's orbital frequency $\Omega$ with the angular frequency of the horizon, the expressions for the two growth rates are seen to be equivalent up to an overall factor associated with differences in the geometry. That these two results are so closely related is not a coincidence, but is a reflection of the fact that a binary and a single black hole both effectively behave like point particles in the long-wavelength limit. Indeed, at leading order in the expansion parameters, the bound (and continuum) states of a Klein--Gordon field on these two spacetimes are mathematically equivalent. All differences between the two cases can therefore be attributed to differences in the corresponding operators $O^L(t)$ that are localized at the origin. (For an EFT approach to single black hole superradiance along these lines, see Ref.~\cite{Endlich:2016jgc}.)

This is not to say that there is nothing novel about orbital superradiance, however. Indeed, the particular ``dumbbell'' geometry of the binary establishes a selection rule that requires $\ell+m$ to be even if the mode is to interact with the binary. Otherwise, the growth rate $\Gamma_\bs$ vanishes. This property can be traced back to the spherical harmonic in \eqref{eq:bound_growth_rate}, or more generally to the spherical harmonics in the matrix elements of \eqref{eq:bound_V}, which are being evaluated with respect to the unit vector $\bmf d$ that is confined to be in the ${z=0}$ plane. As was already pointed out in Ref.~\cite{Wong:2019kru}, the vanishing of these matrix elements has a simple physical interpretation: modes with ${\ell+m \not\in 2\mathbb Z}$ correspond to field profiles that are concentrated away from the ${z=0}$ plane and are therefore unappreciable in the neighborhood of the binary, in which case no interaction can occur. If instead ${\ell+m \in 2\mathbb Z}$, a given mode grows if $0<E_n<m\Omega$ and decays otherwise. 
\vskip-1.5\baselineskip~

\subsection{Mode mixing}
\label{sec:bound_mixing}

Growth rates aside, the geometric properties of a binary also give rise to a number of other interesting effects. From the general solution in \eqref{eq:bound_c1_exp}, it is clear that even if a given mode $\bs$ has zero amplitude ${c_\bs(0) = 0}$ at an initial time ${t=0}$, the presence of another bound state ${\bs' \neq \bs}$ will seed the growth of $\bs$ as long as ${V_{\bs\bs'} \neq 0}$. Likewise, energy from ingoing radiation can also be captured and converted into bound states, as the second line in \eqref{eq:bound_c1_exp} demonstrates. In fact, this is the reason why we have chosen to call the objects in \eqref{eq:bound_V} matrix elements: in analogy with quantum mechanics, they characterize the overlap between different modes as a result of their interaction with the~binary.
\looseness=-1

The appearance of mode mixing is unsurprising in this context, as the underlying spacetime is not time-translation invariant nor is it rotationally symmetric.%
\footnote{Note, however, that the binaries we consider in this paper possess a residual helical symmetry \cite{Friedman:2001pf,*Friedman:2001pf_E} because we have restricted our attention to circular orbits and have, moreover, neglected the emission of gravitational waves.}
Accordingly, the modes ${\bs\equiv(n,\ell,m)}$ and ${\cs\equiv(\w,\ell,m)}$ do not remain eigenstates of the interacting theory. Actually, this last statement needs refining because the matrix elements in \eqref{eq:bound_V} vanish when at least one of ${\ell+m}$ or ${\ell'+m'}$ is odd for reasons already discussed. Thus, modes with $\ell+m\not\in 2\mathbb Z$ are effectively blind to the presence of the black holes and are conserved (at this order in perturbation theory), while modes with $\ell+m\in 2\mathbb Z$ interact with the binary and get mixed into one another.

Let us highlight another consequence of mode mixing: as the solution in \eqref{eq:bound_c1_exp} shows, a given mode $\bs$ will oscillate not just at its natural frequency $E_n$ but also at the secondary frequencies ${|E_n - \Delta_{\bs\bs'}|}$, which results in a beating pattern when viewed in the time domain. Crucially, note that some of these secondary frequencies are much greater than the scalar field's mass~$\mu$. If a particle of the same mass were to have an energy given by one of these high frequencies, we would expect it to escape the gravitational potential of the binary and travel off to infinity. Indeed, the same thing happens in the case of a long-wavelength scalar field, as we show later in Sec.~\ref{sec:cont}.

\subsection{Backreaction and energy extraction}
\label{sec:bound_energy}

\begin{figure*}
\centering\includegraphics{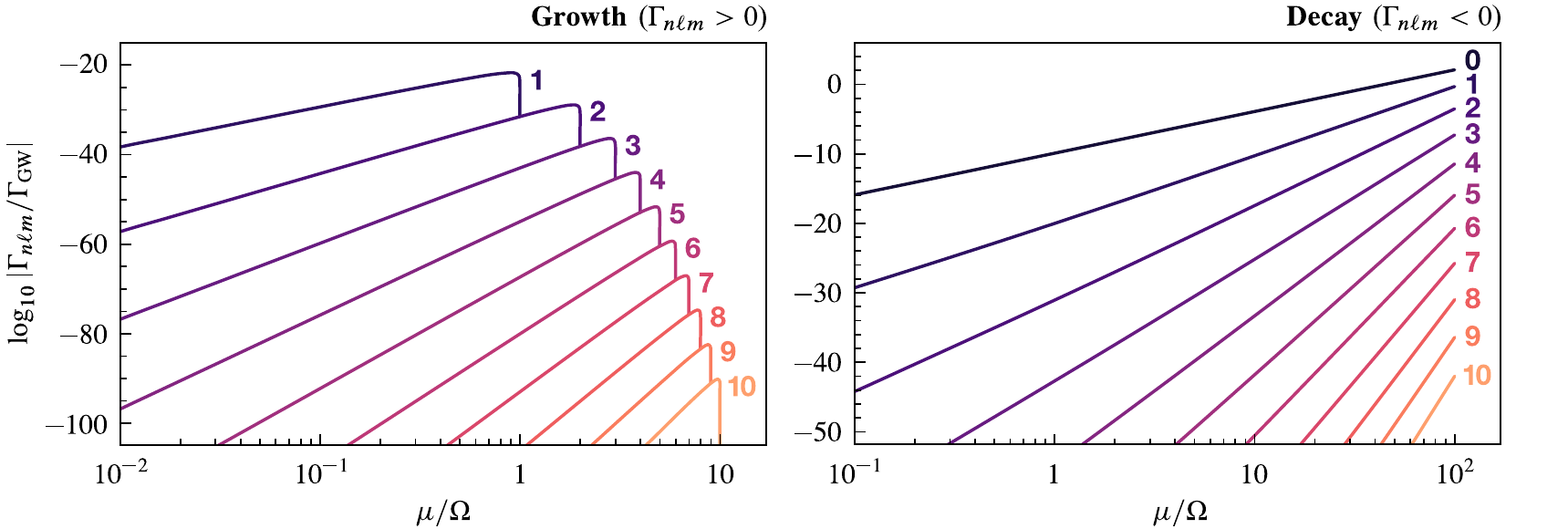}
\caption{The rates $\Gamma_\bs\equiv\Gamma_{n\ell m}$ at which different quasibound states grow or decay around a binary black hole is shown as a function of the scalar field's mass $\mu$ relative to the binary's orbital frequency $\Omega$. They are normalized in units of $\Gamma_\text{GW}$, which is the rate at which the orbit shrinks due to gravitational radiation. The binary itself is taken to be composed of spherical black holes of equal mass traveling with orbital velocity ${v=0.1}$. The growth rates of the ${(n,\ell,m)=(\ell+1,\ell,\ell)}$ modes are shown for ${\ell \in [1,10]}$ in the left panel, whereas the decay rates for the $(\ell+1,\ell,-\ell)$ modes are shown for ${\ell \in [0,10]}$ in the right panel. Note that for a given angular-momentum mode $(\ell,m)$, a larger principal quantum number $n$ would result in a lower rate $\Gamma_{n\ell m}$.}
\label{fig:growth}
\end{figure*}

In many well-motivated scenarios \cite{Arvanitaki:2009fg, Jaeckel:2010ni, Marsh:2015xka,Hui:2016ltb}, an ultralight scalar couples only very weakly to the Standard Model. As such, if a binary black hole is enveloped by a cloud of quasibound states, we should not expect to observe the evolution of this scalar cloud directly and must infer its presence through more indirect means. One possibility is by examining the way it affects the binary's orbital evolution. Because any momentum transferred from the black holes to the scalar field must be accompanied by an appropriate backreaction of the scalar onto the black holes,%
\footnote{This backreaction can be understood in terms of a force that the scalar exerts on the black holes \cite{Wong:2019yoc, Wong:2019kru}.}
any secular increase in the energy $E_b$ of the bound states must have been extracted from the energy stored in the binary.

To gain a sense for how much energy is extracted via this process, let us consider a simplified scenario in which only a single mode ${\hat\bs \equiv (\hat n,\hat\ell, \hat m)}$ is populated initially. Time averaged over a period much longer than the other timescales in the problem, the rate at which energy is extracted from the binary into this bound state is given by \cite{Arvanitaki:2010sy,Yoshino:2013ofa, Brito:2014wla}%
\footnote{At leading order, a number of results from the literature on the superradiant instability of a single Kerr black hole can be adapted to the study of orbital superradiance because, as we mentioned earlier, the mathematics describing the bound states of a Klein--Gordon field in the long-wavelength limit is identical in both cases.}
\bgroup\predisplaypenalty=0
\begin{equation}
\SetDisplaySkip[0][0.6]
	\rule{0pt}{15pt}
	\dot E_b = 2 \Gamma_{\hat\bs} M_{\hat\bs}
\label{eq:bound_energy_rate}
\end{equation}
\egroup
at first order in the interactions, where ${M_{\hat\bs} \coloneq \mu |c^{(0)}_{\hat\bs}|^2}$ is the total (initial) energy in the scalar cloud. Note that the time-averaging procedure eliminates any contribution from mode mixing at this order.

If ${\Gamma_{\hat\bs} > 0}$, the growth of this mode extracts energy from the binary and causes it to inspiral more rapidly than it would in pure vacuum. On the other hand, if ${\Gamma_{\hat\bs} < 0}$, this decaying mode injects energy into the orbit while it is being absorbed and will decelerate the inspiral as a result. Whether either of these effects leave an observable imprint depends on the magnitude of \eqref{eq:bound_energy_rate}. A useful measure is to compare it to the energy flux of gravitational waves emitted by the binary, given to leading order by~\cite{Peters:1963ux}
\begin{equation}
\SetDisplaySkip[0.2]
	\mathcal F = \frac{32}{5} \frac{G^4 M_1^2 M_2^2(M_1 + M_2)}{a^5}
\label{eq:bound_flux_GW}
\end{equation}  
for the case of circular orbits.~Taking the ratio of \eqref{eq:bound_energy_rate} to \eqref{eq:bound_flux_GW}, we find
\begin{equation}
\AddDisplaySkip[-0.6][0.3]
	\frac{ \dot E_b }{ \mathcal F }
	=
	\frac{M_{\hat\bs}}{ E_\text{orb} }
	\frac{2\Gamma_{\hat\bs}}{\Gamma_\text{GW}},
\label{eq:bound_scale_comparison}
\end{equation}
where ${E_\text{orb} = GM_1 M_2/2a}$ is the magnitude of the binary's orbital energy and
\begin{equation}
\AddDisplaySkip[-0.4][0.3]
	\Gamma_\text{GW}
	=
	\frac{64}{5} \frac{G^3 M_1 M_2 (M_1 + M_2)}{a^4}
\label{eq:bound_GW_rate}
\end{equation}
is the rate at which the orbit shrinks due to gravitational radiation.

A cursory glance at \eqref{eq:pert_Bll} and \eqref{eq:bound_growth_rate} will confirm that, for fixed component masses $M_1$ and $M_2$, the scalar cloud's growth rate is largest when both black holes are spherical. In this limit, the ratio $\Gamma_\bs/\Gamma_\text{GW}$ can be expressed as a function of three dimensionless quantities: the binary's orbital velocity $v$, the symmetric mass ratio ${\nu \coloneq M_1 M_2/M^2}$, and the ratio $\mu/\Omega$ that relates the scalar's mass to the binary's orbital frequency. An explicit formula is provided in Appendix~\ref{sec:app_mass_ratio}, where we also argue that the precise value of $\nu$ has little effect on our conclusions. For this reason, we consider only equal-mass binaries in what~follows.

The value of $\Gamma_\bs/\Gamma_\text{GW}$ as a function of $\mu/\Omega$ is shown in Fig.~\ref{fig:growth} for an equal-mass binary composed of spherical black holes. A value of ${v=0.1}$ has been chosen for the orbital velocity, which is large enough that it is  at the limit of validity of the post-Newtonian expansion. As the curves in Fig.~\ref{fig:growth} would all move downwards for smaller values of $v$, they represent the largest-possible rates that we can reliably calculate using this~EFT.

For small values of~$\mu$, the ${\bs=(2,1,1)}$ mode has~the largest growth rate, which reaches a maximum of $\Gamma_{211}/\Gamma_\text{GW}$ ${\sim\! 2 \times 10^{-22}}$ when ${\mu \simeq 9\Omega/10}$. Above the threshold at ${\mu \simeq \Omega}$ (note ${E_{n} \simeq \mu}$), it turns into a decaying mode and leaves the $(3,2,2)$ mode to take over as the fastest-growing mode, until it too becomes a decaying mode at the next threshold ${\mu \simeq 2\Omega}$. This pattern continues for increasing values of $\mu$, with $(\ell+1,\ell,\ell)$ being the fastest-growing mode when ${(\ell-1) \lesssim \mu/\Omega \lesssim \ell}$. The overall trend in Fig.~\ref{fig:growth} clearly shows that the maximum value $\Gamma_\bs$ can attain decreases rapidly as $\mu$ increases. Although the EFT breaks down as we approach ${\mu/\Omega \sim v^{-2}}$, this trend strongly suggests that orbital superradiance is always grossly inefficient. Consequently, the exponential growth of a long-wavelength scalar cloud is unlikely to leave any measurable impact on the evolution of a binary black hole.
\looseness=-1

In contrast, the decay rates can become much larger as $\mu$ increases (see the right panel of Fig.~\ref{fig:growth}), but their observational viability rests on $M_{\hat\bs}$ being comparable to $E_\text{orb}$. The fact that the growth rates are so small implies that clouds with such high densities are unlikely to have formed dynamically around binary black holes that start off in pure vacuum.

Having said that, there may be a possibility that other processes could generate these clouds, particularly in alternative theories of gravity wherein the scalar field is nonminimally coupled to matter. For instance, does the core collapse of a massive star into a black hole remnant leave behind an appreciable scalar cloud? If so, could successive supernova events in a stellar binary lead to a pair of black holes enveloped by a common scalar cloud (assuming an optimal value for $\mu$)? It has been shown that a large amount of scalar radiation can be produced during core collapse in a certain class of scalar--tensor theories \cite{Gerosa:2016fri, Sperhake:2017itk, Rosca-Mead:2019seq, Cheong:2018gzn}, although current numerical methods are unable to determine if a scalar cloud can develop around a black hole remnant \cite{PC_Sperhake}. Exploring these open questions presents an exciting opportunity for future work.

\section{Outgoing radiation}
\label{sec:cont}

The periodic forcing that a binary exerts on a surrounding cloud inevitably leads to a fraction of the scalar field being upscattered and ejected as outgoing radiation. Additionally, ingoing radiation can scatter off this binary and undergo amplification when given the right initial conditions. In our perturbative approach, both of these phenomena are encoded in the principal value of the $\w$ integral in \eqref{eq:pert_PI_cont_exp}. Using \eqref{eq:app_R_analytic_cont} to replace the Whittaker function with the radial solution $R^+_\ell(k,r)$, the result is
\begin{equation}
	\phi^{(1)}_c(x)
	=
	\sum_{\ell,m}\int_{\w \in \mathbb R\backslash\{\pm E_n\}} \frac{\dx\w}{2\pi}
	\mathcal A^{(1)}_\cs \phi^+_{k\ell m}(\bmf x) e^{-i\w t},
\label{eq:cont_phi_1}
\end{equation}
where the first-order correction to the outgoing amplitude for a given mode ${\cs\equiv(\w,\ell,m)}$ is
\begin{subequations}
\label{eq:cont_A_main}
\begin{align}
	\mathcal A^{(1)}_\cs
	&=
	s_\ell(\zeta) k^{\ell+1}
	\int\dx t\, O_{\ell m}^{(0)}(t) e^{i\w t},
	\label{eq:cont_A}
	\\
	s_\ell(\zeta)
	&=
	\frac{\Gamma(\ell+1+i\zeta)}{\Gamma(\ell+1)}
	e^{-\pi\zeta/2-i\sigma_\ell(\zeta)}.
	\label{eq:cont_s}
\end{align}
\end{subequations}

It is worth highlighting that the solution in \eqref{eq:cont_phi_1} is not of the form ${\phi = \phi^> + \text{c.c.}}$; hence, $\mathcal A^{(1)}_\cs$ implicitly accounts for both positive- and negative-frequency modes. Because the overall solution must be real, these coefficients will have to satisfy a constraint analogous to \eqref{eq:pert_Olm_constraint}. Combining \eqref{eq:app_cc_phi_+} with the freedom to relabel ${\w \to -\w}$ and ${m \to -m}$ as they are being integrated and summed over, respectively, we arrive at the constraint
\begin{equation}
\AddDisplaySkip[-0.3][0.4]
	\mathcal A^{(1)}_{\w\ell m} = e^{-\pi\zeta} (-1)^{\ell+m} \mathcal A^{(1)*}_{-\w\ell-m}.
\label{eq:cont_A_constraint}
\end{equation}
One can now verify that the solution in \eqref{eq:cont_A_main} adheres to this constraint after using \eqref{eq:pert_Olm_constraint} along with the identities ${ [\Gamma(z)]^* \equiv \Gamma(z^*) }$ and ${ k^*(\w) \equiv -k(-\w) }$; the latter following directly from the definition in \eqref{eq:pert_k}.

To compute the rate $\dot E_\text{SW}$ at which energy is carried away from the binary in the form of scalar waves, it is useful to first recast the zeroth-order solution \eqref{eq:pert_phi_0_short} into a form similar to \eqref{eq:cont_phi_1}. The identity in \eqref{eq:app_cc_phi_0} can be used to rewrite the sum over continuum states as
\begin{equation}
\AddDisplaySkip[0.3][-0.1]
	\phi_c^{(0)}(x) = \sum_\cs
	2\mathcal I_\cs \phi_{k\ell m}(\bmf x)
	e^{-i\w t},
\label{eq:cont_phi_0_ingoing_only}
\end{equation}
where
$\mathcal I_\cs \equiv \mathcal I_{\w\ell m} = \mathcal I^>_{\w\ell m} + e^{\pi\zeta}(-1)^{\ell+m}\mathcal I^{>*}_{-\w\ell-m}$
implicitly accounts for the ingoing amplitudes of both positive- and negative-frequency modes, and can be seen to satisfy the requisite constraint
\begin{equation}
	\mathcal I_{\w\ell m} = e^{\pi\zeta} (-1)^{\ell+m} \mathcal I^*_{-\w\ell-m}.
\label{eq:cont_I_constraint}
\end{equation}
Taken in combination with \eqref{eq:cont_phi_1}, the full solution
$\phi = \phi^{(0)} + \phi^{(1)} + \cdots$
for the continuum states has the form
\begin{equation}
\AddDisplaySkip[0.3][-0.1]
	\phi_c(x) = \sum_\cs
	[ \mathcal I_\cs \phi^-_{k\ell m}(\bmf x) e^{-i\w t}
	+
	\mathcal R_\cs \phi^+_{k\ell m}(\bmf x) e^{-i\w t}],
\end{equation}
where $\mathcal R_\cs \coloneq \mathcal I_\cs + \mathcal A_\cs$ is the total outgoing amplitude, with $\mathcal A_\cs$ given to first order in the interactions in \eqref{eq:cont_A_main}.

Now integrating the $(t,r)$ component of the scalar's energy--momentum tensor over a spherical shell of radius $r$ and taking the limit ${r \to \infty}$, the time-averaged power loss is given by the difference between the energy flux flowing into and out of the system,
${ \dot E_\text{SW}^{\vphantom{\text{out}}} = \dot E_\text{SW}^\text{out} - \dot E_\text{SW}^\text{in} }$.
These quantities have simple expressions when integrated over all time:
\begin{subequations}
\label{eq:cont_dot_E}
\begin{align}
	\int_{-\infty}^\infty \dx t\, \dot E_\text{SW}^\text{out}
	&=
	\sum_\cs \theta(k^2) \frac{\w}{k} | \mathcal R_\cs |^2,
	\label{eq:cont_dot_E_out}
	\allowdisplaybreaks\\
	\int_{-\infty}^\infty \dx t\, \dot E_\text{SW}^\text{in}
	&=
	\sum_\cs \theta(k^2) \frac{\w}{k} | \mathcal I_\cs |^2.
	\label{eq:cont_dot_E_in}
\end{align}
\end{subequations}

\subsection{Ejection of bound states}
\label{sec:cont_ejection} 

To better understand the physical implications of \eqref{eq:cont_phi_1}, let us start by supposing---as we did in Sec.~\ref{sec:bound_energy}---that there is no ingoing radiation and only a single bound state $\hat\bs$ is populated at zeroth order. In this case, the outgoing radiation we compute represents the portion of the scalar cloud that is being ejected out of the system. The energy flux for this process is given to leading order by
\begin{align}
	\dot E_\text{SW}
	&=
	\frac{1}{2\pi\delta(0)} \sum_\cs \theta(k^2) \frac{\w}{k} |\mathcal A^{(1)}_\cs |^2,
\label{eq:cont_energy_rate_ejection}
\end{align}
where the delta function in the denominator is associated with the integral over all time, ${\int\dx t \equiv 2\pi\delta(0)}$. Because this formula is quadratic in $\mathcal A^{(1)}_\cs$, it is formally of second order in the interactions, and thus $\dot E_\text{SW}$ will generally be smaller than the rate $\dot E_b$ at which energy extracted from the binary fuels the growth of bound states; cf.~\eqref{eq:bound_energy_rate}. However, this hierarchy becomes inverted when the scalar field is sufficiently light, as we will show.

Since $\zeta$ and, consequently, $\sigma_\ell(\zeta)$ are real when ${k^2 > 0}$ \cite{DLMF}, taking the absolute square of \eqref{eq:cont_A_main} yields
\begin{align}
	\big| \mathcal A^{(1)}_\cs \big|^2
	&=
	S_\ell(\zeta) k^{2(\ell+1)}
	\left| \int\dx t\, O^{(0)}_{\ell m}(t) e^{i\w t} \right|^2.
\label{eq:cont_ejection_A2}
\end{align}
One might recognize ${S_\ell(\zeta) \equiv |s_\ell(\zeta)|^2}$ as the Sommerfeld enhancement factor~\cite{Iengo:2009ni, Cassel:2009wt, ArkaniHamed:2008qn}, which may be rewritten as
\begin{equation}
	S_\ell(\zeta) = \frac{1}{(\ell!)^2}
	\frac{\pi\zeta e^{-\pi\zeta}}{\sinh(\pi\zeta)}
	 \prod_{j=1}^\ell (j^2 + \zeta^2)
\label{eq:cont_sommerfeld}
\end{equation}
after using standard identities for the gamma function \cite{Iengo:2009ni}. On the mathematical level, this factor arises naturally in our calculations because the interaction terms involve evaluating derivatives of Coulomb functions at the origin. We run into difficulties, however, when attempting to assign to this factor its usual physical interpretation. We elaborate further in later parts of this section.

When only a single bound state $\hat\bs$ is populated, a close inspection of \eqref{eq:pert_Olm_explicit} reveals that mode mixing will generate continuum states with frequencies in a discrete set given by $\w_m = |E_{\hat n} + (m-\hat m)\Omega|$. As we argued at the end of Sec.~\ref{sec:bound_mixing}, these continuum states are generated alongside newly populated quasibound states ${u \neq \hat u}$, which oscillate at the same set of frequencies $|E_n - \Delta_{u\hat u}| \equiv |E_{\hat n} + (m-\hat m)\Omega|$. The subset of these continuum states with ${\w_m > \mu}$ are radiation modes that propagate to infinity. Substituting \eqref{eq:pert_Olm_explicit} into \eqref{eq:cont_energy_rate_ejection}, the power in these radiation modes is
\begin{align}
	\dot E_\text{SW}
	&=
	\frac{1}{2\mu}\sum_{\ell,m} 
	\left|
		\frac{Y^*_{\ell m}(\bmf d) Y_{\hat\ell\hat m}(\bmf d)}{(2\ell+1)!!}
		B_{\ell\hat\ell} \mathfrak R_{\hat n\hat\ell} 
		(\hat m\Omega - E_{\hat n}) c_{\hat\bs}^{(0)}
	\right|^2
	\nonumber\\
	&\quad\times
	\theta(k_m^2) [S_\ell(\zeta_m) + S_\ell(-\zeta_m)]
	\w_m k_m^{2\ell+1},
\label{eq:cont_ejection_E_rad}
\end{align}
where ${k_m \equiv k(\w_m)}$ and likewise ${\zeta_m \equiv \zeta(\w_m)}$. (The intermediate steps for this general type of calculation are presented in Appendix~\ref{sec:app_flux}.)

As it stands, this result is not particularly illuminating. It is perhaps most instructive to compare \eqref{eq:cont_ejection_E_rad} to \eqref{eq:bound_energy_rate}, in which case one finds
\begin{align}
	\frac{ \dot E_\text{SW} }{ \dot E_b }
	&=
	\sum_{\ell,m} 
	\left| \frac{Y_{\ell m}(\bmf d)}{(2\ell+1)!!} \right|^2
	\frac{ E_{\hat n} B_{\ell\hat\ell}^2}{2\mu^2 B_{\hat\ell\hat\ell}}
	(\hat m\Omega - E_{\hat n})
	\nonumber\\
	&\quad\times
	\theta(k_m^2) [S_\ell(\zeta_m) + S_\ell(-\zeta_m)]
	\w_m k_m^{2\ell+1}.
\label{eq:cont_ejection_E_ratio}
\end{align}
Although this is a sum over infinitely many modes, it suffices to keep only the lowest few values of~$\ell$ to obtain a good approximation because the prefactor $|Y_{\ell m}(\bmf d)/(2\ell+1)!!|^2$ decays rapidly like $\sim 1/(2\ell+1)!$.  To proceed, let us begin by analyzing the limiting behavior of a given term in \eqref{eq:cont_ejection_E_ratio} when ${k_m^2 > 0}$ and ${\zeta_m \gg 1}$. The sum of Sommerfeld factors reads%
\footnote{This sum is even in $\zeta$ and ensures that physical results are independent of our choice of sign for $\w$.}
\begin{equation}
	S_\ell(\zeta) + S_\ell(-\zeta)
	=
	\frac{1}{(\ell!)^2}
	2\pi\zeta \coth(\pi\zeta)\prod_{j=1}^\ell (j^2 + \zeta^2),
\label{eq:cont_symmetric_sum_Sommerfeld}
\end{equation}
which has the asymptotic form $\sim2\pi |\zeta|^{2\ell+1}$ when ${\zeta\to \infty}$. While it may be natural to want to think of this as describing the usual Sommerfeld enhancement for low-momentum modes (recall ${\zeta \propto 1/k}$), in the present context there is no analogous process that occurs on flat space, since bound states cannot form in the absence of the binary's gravitational potential. With this in mind, the Sommerfeld factors appearing in \eqref{eq:cont_ejection_E_ratio} are perhaps best regarded as simply an inevitable part of the result rather than engendering any kind of enhancement.

Taking the limit ${\zeta_m \gg 1}$, the corresponding term in \eqref{eq:cont_ejection_E_ratio} reduces to
\begin{equation}
	\left| \frac{Y_{\ell m}(\bmf d)}{(2\ell+1)!!} \right|^2
	\frac{ E_{\hat n} B_{\ell\hat\ell}^2}{2\mu B_{\hat\ell\hat\ell}}
	2\pi (GM \mu^2)^{2\ell+1}
	(\hat m\Omega - E_{\hat n}),
\label{eq:cont_ejection_low_momentum}
\end{equation}
since ${\w_m \simeq \mu}$ in this case. Power counting reveals that this term scales with the EFT's expansion parameters as
${ \sim v^5 (a/\lambda_\text{dB})^{2\ell+1} (\hat m - \mu/\Omega) }$.
Accordingly, for clouds with ${\hat m \sim\mathcal O(1)}$ and ${\mu/\Omega \ll v^{-2}}$ [cf.~\eqref{eq:eft_cutoff_mu}], the rate at which energy is carried away by low-momentum radiation is parametrically suppressed relative to $\dot E_b$.

Indeed, each term in \eqref{eq:cont_ejection_E_ratio} is a monotonically increasing function of $k_m$, so most of the energy is carried away in high-momentum modes (${\zeta_m \ll 1}$). In this limit, ${S_\ell(\zeta) \sim 1}$, and thus the corresponding term in \eqref{eq:cont_ejection_E_ratio} reduces to
\begin{equation}
	\left| \frac{Y_{\ell m}(\bmf d)}{(2\ell+1)!!} \right|^2
	\frac{ E_{\hat n} B_{\ell\hat\ell}^2}{\mu^2 B_{\hat\ell\hat\ell}}
	\w_m k_m^{2\ell+1} (\hat m\Omega - E_{\hat n}).
\label{eq:cont_ejection_high_momentum}
\end{equation}
To assess the typical size of this term, it is instructive to express it in terms of the ratio ${f_\mu \coloneq \mu/\Omega}$. Also using the definition ${ \w_m = | E_{\hat n} + (m-\hat m)\Omega| }$ and approximating ${ E_{\hat n} \simeq \mu }$, \eqref{eq:cont_ejection_high_momentum} becomes
\begin{align}
	&	
	\frac{1}{f_\mu} \left| \frac{Y_{\ell m}(\bmf d)}{(2\ell+1)!!} \right|^2
	\frac{B_{\ell\hat\ell}^2}{B_{\hat\ell\hat\ell}}
	\Omega^{2\ell+2}
	(\hat m - f_\mu)(f_\mu + \Delta m)
	\nonumber\\&\times
	(2 f_\mu \Delta m + \Delta m^2)^{\ell+1/2},
\label{eq:cont_ejection_high_momentum_f}
\end{align}
where ${\Delta m = m-\hat m}$. This term scales as $\sim v^{2\ell+6}$ when $f_\mu$, $\hat m$, and $\Delta m$ are all of order unity, meaning the rate at which energy is carried away by high-momentum radiation is---in this case---also parametrically suppressed relative to $\dot E_b$. However, when ${f_\mu \ll 1}$, the $1/f_\mu$ prefactor enhances this term such that power loss to radiation can become significant in comparison to $\dot E_b$. More precisely, a given high-momentum mode will extract energy from the cloud at a rate greater than $\dot E_b$ if%
\footnote{This is a conservative upper bound that does not take the possibility that $B^2_{\ell\hat\ell}$ can vanish into consideration. As an example, if an equal-mass binary is surrounded by a cloud comprised of only the ${\hat\bs=(2,1,1)}$ mode, one finds ${B_{01}=0}$; hence, energy is predominantly radiated away in the ${\ell=1}$ modes. In this case, $\dot E_\text{SW}$ becomes larger than $\dot E_b$ only when $\mu/\Omega \lesssim v^8$.}
\begin{equation}
	\mu/\Omega \lesssim v^{2\ell+6} \leq v^6.
\end{equation}

One may conclude from this simple scaling analysis that scalar clouds cannot form dynamically around binary black holes when the scalar field's mass $\mu$ is sufficiently light, as the rate at which the cloud is depleted via scalar radiation is greater than the rate at which it grows due to orbital superradiance. As a rough guide, this occurs when
$\mu \lesssim 10^{-19}~\text{eV} \, (v/0.1)^3(M_\odot/M)$.

\paragraph{Gravitational waves}
It is worth briefly remarking that a scalar cloud will also emit gravitational waves due to the oscillatory nature of its backreaction onto the spacetime. When only the single bound state $\hat\bs$ is populated (and further assuming ${\hat\ell=\hat m}$ for simplicity), the energy flux of gravitational waves emitted by the cloud is~\cite{Yoshino:2013ofa,Brito:2014wla}
\begin{equation}
	\dot E_\text{GW}
	\simeq
	\frac{\mathcal C_{\hat\bs}}{G} \left( \frac{M_{\hat\bs}}{M}\right)^2 (GM\mu)^{4\hat\ell+10},
\end{equation}
where $\mathcal C_{\hat\bs}$ ($<1$) is some dimensionless prefactor whose exact form will not be important to us, but we note that ${\mathcal C_{n00} = 0}$. Comparing this with \eqref{eq:bound_energy_rate}, we find
${ \dot E_\text{GW}/\dot E_b \sim (M_{\hat\bs}/M) f_\mu^4 v^{4\hat\ell+9} }$. We should expect ${M_{\hat\bs}/M < 1}$ if the scalar is to behave like a test field around the binary, and thus $\dot E_\text{GW}$ is generally much smaller than $\dot E_b$. Moreover, unless ${\hat\ell=1}$ and $f_\mu$ is close to the UV cutoff of this EFT [cf.~\eqref{eq:eft_cutoff_mu}], a comparison with \eqref{eq:cont_ejection_high_momentum_f} reveals that $\dot E_\text{GW}$ is also typically smaller than~$\dot E_\text{SW}$.

\subsection{Superradiant scattering}

We now turn our attention to a different setup in which the zeroth-order solution is given by a steady stream of radiation. In realistic astrophysical scenarios, we would expect an incident wave to be essentially planar on the scales of the binary, but such a configuration turns out to be difficult to analyze in the present context. Specifically, because a plane wave can always be written as a linear combination of spherical waves, this zeroth-order solution will contain infinitely many modes that subsequently mix into one another as a result of their interactions with the binary. To render the following discussion more tractable, we will consider a simpler, albeit more artificial setup that should nevertheless suffice for illustrating the most salient features. The more realistic case of plane waves is left for future work.

With this in mind, let us consider a steady stream of ingoing radiation peaked at the single mode ${\hat\cs\equiv(\hat\w,\hat\ell,\hat m)}$. This corresponds to making the choice
\begin{equation}
	\mathcal I^>_{\w\ell m} = \Phi_{\hat\cs} 2\pi\delta(\w-\hat\w)
	\delta_{\ell\hat\ell}\delta_{m\hat m},
\label{eq:cont_single_ingoing_rad_mode}
\end{equation}
where $\Phi_{\hat\cs}$ is in general some complex-valued coefficient with dimensions of energy. It has previously been shown that this ingoing wave can extract energy from the binary's orbital motion and undergo amplification under the right conditions \cite{Wong:2019kru}. In this subsection, we extend the results of Ref.~\cite{Wong:2019kru} in several directions.

\paragraph{Amplification factor}
From \eqref{eq:pert_Olm_explicit}, we learn that a single ingoing mode $\hat\cs$ will scatter into multiple outgoing modes with frequencies in a discrete set given by ${\w_m = |\hat\w + (m-\hat m)\Omega|}$. Included in this spectrum is the original (or ``primary'') mode with frequency ${\hat\w\equiv \w_{\hat m}}$, which typically comprises the majority of the outgoing energy flux as a result of its interference with the ingoing radiation. Explicitly, one expands \eqref{eq:cont_dot_E} to find
\begin{equation}
	\int\dx t\,\dot E_\text{SW}
	=
	\sum_\cs \theta(k^2) \frac{\w}{k}
	2\Re \mathcal I^*_\cs \mathcal A^{(1)}_\cs
	+
	\mathcal O(\mathcal A^2).
\end{equation}
Now substituting \eqref{eq:pert_Olm_explicit} into the above formula and making use of the symmetries in \eqref{eq:cont_A_constraint} and \eqref{eq:cont_I_constraint}, we obtain
\begin{equation}
	\dot E_\text{SW}
	=
	2\Re
	\frac{|2Y_{\hat\ell\hat m}(\bmf d)|^2}{(2\hat\ell+1)!!}
	B_{\hat\ell\hat\ell}
	|\Phi_{\hat\cs}|^2
	s_{\hat\ell}(\hat\zeta) C_{\hat\ell}(\hat\zeta)
	\hat\w\hat k^{2\hat\ell} (\hat m\Omega - \hat\w)
\end{equation}
to first order in the interactions.~(The details of this calculation are presented in Appendix~\ref{sec:app_flux}.) As the Gamow factor
$ C_{\ell}(\zeta) \equiv s_{\ell}^*(\zeta)/(2\ell+1)!! $
when ${\zeta\in\mathbb R}$, the above expression is already real and can be further simplified to read
\begin{equation}
	\dot E_\text{SW} =
	\left|\frac{2Y_{\hat\ell\hat m}(\bmf d)}{(2\hat\ell+1)!!}\right|^2
	2B_{\hat\ell\hat\ell} |\Phi_{\hat\cs}|^2
	S_{\hat\ell}(\hat\zeta) \hat\w\hat k^{2\hat\ell} (\hat m\Omega - \hat\w).
\label{eq:cont_scattering_E_rad_1}
\AddDisplaySkip[-0.2][0.2]
\end{equation}

\begin{figure*}
\centering\includegraphics{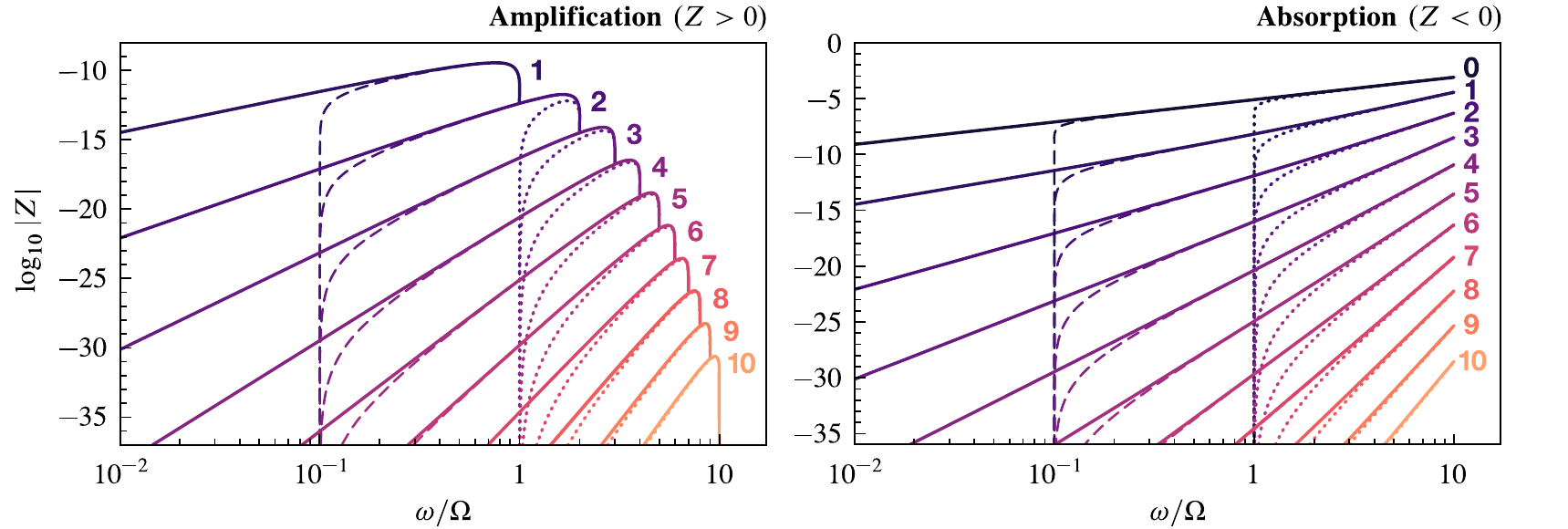}
\caption{The amplification factor $Z$ for a single ingoing radiation mode is shown as a function of its frequency $\w$ in units of the binary's orbital frequency $\Omega$. The binary itself is taken to be composed of spherical black holes of equal mass traveling with orbital velocity ${v=0.1}$. The values of $Z$ for the ${\ell=m}$ modes are shown for ${\ell \in [1,10]}$ in the left panel, while in the right panel we plot the corresponding values for the ${\ell=-m}$ modes in the range ${\ell \in [0,10]}$. In both panels, the amplification factors are shown for three different values of the scalar field's mass: $\mu=0$~(solid lines), $\mu=0.1\Omega$ (dashed lines), and $\mu=\Omega$ (dotted lines).}
\label{fig:Z}
\end{figure*}

To gain a sense of how much energy is exchanged during this scattering process, it is instructive to compare \eqref{eq:cont_scattering_E_rad_1} to the total flux of ingoing radiation $\dot E_\text{SW}^\text{in}$, an expression for which is also derived in Appendix~\ref{sec:app_flux} [cf.~\eqref{eq:app_energy_ingoing_flux}]. This dimensionless ratio defines the total amplification factor ${Z \coloneq \dot E_\text{SW}/\dot E_\text{SW}^\text{in} }$, which is given by
\bgroup\predisplaypenalty=0
\begin{equation}
	Z =
	\bigg( \frac{2S_{\hat\ell}(\hat\zeta)}{1 + e^{-2\pi\hat\zeta}} \bigg)
	\bigg| \frac{2Y_{\hat\ell\hat m}(\bmf d)}{(2\ell+1)!!} \bigg|^2
	B_{\hat\ell\hat\ell}
	\hat k^{2\hat\ell+1} (\hat m\Omega - \hat\w)
\label{eq:cont_Z}
\AddDisplaySkip[][0.2]
\end{equation}
\egroup
in the case of a single ingoing mode $\hat\cs$. As was the case for the growth rate $\Gamma_\bs$, this result is identical---up to a geometric factor and appropriate identifications of $\Omega$ and~$M$---to the amplification factor for a long-wavelength scalar field scattering off a single rotating black hole \cite{Starobinsky:1973aij, Brito:2015oca, Endlich:2016jgc}. Note that the comparison can only be made for massless scalars, however, as an analytic expression for the amplification of massive scalar waves by a Kerr black hole is not presently known (to the best of our knowledge).

Returning to the result in \eqref{eq:cont_Z}, observe that a key feature in the massive case is the appearance of the Sommerfeld factor~$S_{\hat\ell}$. For high-momentum modes with ${\hat k \gg GM\mu^2}$ (${\hat\zeta \ll 1}$), $S_{\hat\ell}(\hat\zeta)$ and the exponential $e^{-2\pi\hat\zeta}$ both reduce to unity such that we recover the result in Ref.~\cite{Wong:2019kru}. This limiting behavior signifies that the binary's long-range gravitational potential has negligible influence on the amplification or absorption of high-momentum modes. For more general values of $\hat k$, \eqref{eq:cont_sommerfeld} may be used to show that
\begin{equation}
\AddDisplaySkip[-0.3]
	\frac{2S_{\ell}(\zeta)}{1 + e^{-2\pi\zeta}}
	=
	\frac{1}{(\ell!)^2}
	\frac{2\pi\zeta}{\sinh(2\pi\zeta)}\prod_{j=1}^\ell (j^2 + \zeta^2),
\label{eq:cont_Sommerfeld_suppression}
\end{equation}
which has the asymptotic form ${\sim |\zeta|^{2\ell+1} e^{-2\pi|\zeta|}}$ when $\zeta\to\infty$. This leads to an exponential suppression of the energy carried away by low-momentum modes.

Interestingly, one might naively expect that the appearance of $S_\ell(\zeta)$ in \eqref{eq:cont_scattering_E_rad_1} should lead to Sommerfeld enhancement, but this is directly contradicted by the result in \eqref{eq:cont_Sommerfeld_suppression}, which we interpret as follows. In the classical analog of Sommerfeld enhancement, we imagine a stream of particles impinging on a star of radius $R_\star$ \cite{ArkaniHamed:2008qn}.~In the absence of gravity, the geometric cross section ${\sigma_0 = \pi R_\star^2}$ of this star provides a measure of the fraction of particles that collide with it and are subsequently absorbed. However, the actual cross section~$\sigma$ for this interaction can be much larger, especially for particles with low momenta, because the star's attractive gravitational potential is able to pull in particles that have impact parameters greater than~$R_\star$.

With this picture in mind, one should now expect no analogous enhancement to occur in the present scenario. In our setup, the ingoing mode $\hat\cs$ is a spherical wave that is already directed straight at the origin; hence, the presence of the binary's gravitational potential does nothing to affect the amount of radiation that reaches it. While this argues for the lack of Sommerfeld enhancement, it remains to explain the suppression of low-momentum modes observed in \eqref{eq:cont_Sommerfeld_suppression}. Although the physical origin of this suppression is still not fully understood, the most likely explanation is that it is due to the conversion of radiation modes into bound states [cf.~the second line in \eqref{eq:bound_c1_exp}], which is enhanced at low momenta. A full quantitative analysis is needed to validate this interpretation, although such a task is beyond the scope of this present paper. 

Putting these conceptual issues aside, let us discuss the likelihood of observing this energy exchange between the binary and the scalar. As we discussed earlier in Sec.~\ref{sec:bound}, the particular geometry of the binary prevents it from interacting with any long-wavelength mode whose angular momentum is such that ${\ell+m \not\in2\mathbb Z}$. For the remaining modes, amplification occurs if ${0 < \hat\w < \hat m \Omega}$, in which case the binary loses energy and inspirals more rapidly as a result. Otherwise, there is a net absorption of the scalar by the binary, which then gains energy and experiences a slowing down of its inspiral. The feasibility of observing either of these effects depends on the magnitude of $\dot E_\text{SW}$ when compared to the outgoing flux of gravitational radiation $\mathcal F$. As a rough estimate, we should expect to observe the influence of this energy exchange on the orbital motion only if the ratio
\begin{equation}
\AddDisplaySkip[-0.4][0.4]
	\frac{\dot E_\text{SW}}{\mathcal F}
	=
	Z \frac{ \dot E_\text{SW}^\text{in} }{\mathcal F}
\end{equation}
is not too much smaller than unity.

The amplification factor $Z$ is shown as a function of the ingoing frequency for different values of the scalar field's mass in Fig.~\ref{fig:Z}. As we did in Sec.~\ref{sec:bound_energy}, we assume an equal-mass binary composed of spherical black holes traveling with orbital velocity ${v=0.1}$. The same reasoning as before justifies limiting ourselves to this specific case. First, the curves in Fig.~\ref{fig:Z} would all move downwards for smaller values of $v$, so once again they represent the largest possible values that can be reliably calculated using this EFT. Moreover, the precise value of the binary's mass ratio has little effect on our overall conclusions, as Appendix~\ref{sec:app_mass_ratio} argues.

In the left panel of Fig.~\ref{fig:Z}, we see that amplification is most pronounced for the ${\ell=m=1}$ mode, which reaches a maximum value ${Z \sim 4\times 10^{-10}}$ when ${\w = 3\Omega/4}$. As the higher ${\ell= m}$ modes are less efficiently amplified, the overall trend suggests that this orbital superradiant mechanism continues to become increasingly insignificant even for large frequencies ${\w \gtrsim \Omega/v}$ beyond the EFT's regime of validity. Given the smallness of $Z$ and the unlikelihood that the ingoing flux $\dot E_\text{SW}^\text{in}$ of scalar waves would match or exceed the outgoing gravitational-wave flux $\mathcal F$ in realistic astrophysical scenarios, we deduce that the amplification of long-wavelength scalar fields is observationally inaccessible. Granted, this conclusion is based on a rather artificial setup, although it seems unlikely to change were we to consider the more realistic case of plane~waves.

In contrast, the right panel of Fig.~\ref{fig:Z} demonstrates that absorption continues to become more efficient as $\w$ increases; naturally prompting us to ask: is there any regime (possibly at some frequency $\w \gtrsim \Omega/v$ outside the EFT's regime of validity) in which $|Z|$ and $\dot E_\text{SW}^\text{in}$ are both large enough that they can leave a measurable imprint on the evolution of the binary? On a more theoretical level, it is also interesting to ask: what is the maximum amount of radiation that can be absorbed by a binary black hole in a given time? Because the amplification factor is bounded from below (${Z \geq -1}$), there are two possibilities for what might occur in the high-frequency regime: either $Z$ gradually tends to a minimum value (meaning absorption would be most pronounced at high frequencies), or it has a turning point (i.e., there is a critical frequency beyond which absorption becomes less efficient again). The fact that moving black holes can amplify high-frequency radiation \cite{Cardoso:2019dte} (via what is essentially the slingshot effect) is a hint that the latter may be more likely. These questions point to potential directions for future work.

\paragraph{Secondary modes}
To complete our discussion on the scattering of scalar waves, we ought to discuss the additional energy that is carried away by the secondary modes ${\cs\neq\hat\cs}$ generated through mode mixing. Although their contribution to the energy flux is typically subleading because they first appear in $\dot E_\text{SW}$ at second order in the interactions, the energy they carry can exceed that of the primary mode if $\hat\w$ is sufficiently small, as we now show.

Let us denote this $\mathcal O(\mathcal A^2)$ correction to the energy flux as $\dot E^{(2)}_\text{SW}$. The calculation is almost identical to that in Sec.~\ref{sec:cont_ejection} and the end result is found to be
\begin{align}
	&\qquad\quad
	\dot E^{(2)}_\text{SW}
	=
	\frac{1}{2\pi\delta(0)} \sum_\cs \theta(k^2) \frac{\w}{k}
	\big| \mathcal A^{(1)}_\cs \big|^2
	\label{eq:cont_scattering_2_dot_E}
	\allowdisplaybreaks\\
	&=
	\sum_{\ell,m}
	\left|
		\frac{2\Phi_{\hat\cs} Y^*_{\ell m}(\bmf d) Y_{\hat\ell\hat m}(\bmf d)}%
		{(2\ell+1)!! (2\hat\ell+1)!!}
	\right|^2
	B_{\ell\hat\ell}^2 S_{\hat\ell}(\hat\zeta) \hat k^{2\hat\ell}
	(\hat m\Omega - \hat\w)^2
	\nonumber\\
	&\quad\times
	\theta(k_m^2) [S_\ell(\zeta_m) + S_\ell(-\zeta_m)]
	\w_m k_m^{2\ell+1},
\end{align}
where the frequencies of the modes being summed over are given by ${\w_m = |\hat\w + (m-\hat m)\Omega|}$. To see that this second-order correction can be much larger than the outgoing flux at $\mathcal O(\mathcal A)$ in \eqref{eq:cont_scattering_E_rad_1}, which we here denote by $\dot E^{(1)}_\text{SW}$, we simply divide one by the other to find
\begin{align}
	\frac{ \dot E^{(2)}_\text{SW} }{ \dot E^{(1)}_\text{SW} }
	&=
	\sum_{\ell,m}
	\frac{1}{\hat\w}
	\left|
		\frac{Y_{\ell m}(\bmf d)}{(2\ell+1)!!}
	\right|^2
	\frac{B^2_{\ell\hat\ell}}{ B_{\hat\ell\hat\ell} }
	(\hat m\Omega - \hat\w)
	\nonumber\\&\quad\times
	\theta(k_m^2) [S_\ell(\zeta_m) + S_\ell(-\zeta_m)]
	\w_m k_m^{2\ell+1}.
\label{eq:cont_scattering_2_1_ratio}
\end{align}

Observe that \eqref{eq:cont_scattering_2_1_ratio} has the same mathematical structure as \eqref{eq:cont_ejection_E_ratio}; hence, the analysis will proceed in a largely similar fashion. First, recall that while sums of this kind are to be taken over infinitely many modes, a good approximation can be obtained by keeping only the lowest few values of $\ell$, since the higher multipoles are factorially suppressed. Next, the fact that each term in \eqref{eq:cont_scattering_2_1_ratio} is a monotonically increasing function of $k_m$ signifies that most of the energy carried away will be in the form of high-momentum modes ${(\zeta_m \ll 1)}$.

Therefore, let us concentrate on a given term in \eqref{eq:cont_scattering_2_1_ratio} and suppose that ${\zeta_m \ll 1}$. In terms of the dimensionless ratio ${f_{\hat\w}\coloneq\hat\w/\Omega}$, this term reads
\begin{align}
	&
	\frac{1}{f_{\hat\w}}
	\left| \frac{Y_{\ell m}(\bmf d)}{(2\ell+1)!!} \right|^2
	\frac{B^2_{\ell\hat\ell}}{B_{\hat\ell\hat\ell} }
	\Omega^{2\ell+2}
	(\hat m - f_{\hat\w})(\Delta m + f_{\hat\w})
	\nonumber\\&\times
	[(\Delta m + f_{\hat\w})^2 - (\mu/\Omega)^2]^{\ell+1/2}
\end{align}
where ${\Delta m = m - \hat m}$. When $f_{\hat\w}$, $\hat m$, and $\Delta m$ are all of order unity, this term scales as ${\sim v^{2\ell+6}}$ and is thus parametrically suppressed. However, if instead ${f_{\hat\w} \ll 1}$ (and necessarily ${\mu/\Omega < f_{\hat\w}}$ if $\hat\cs$ is to be a radiation mode), this term can become arbitrarily large. As a result, the energy carried away in a secondary mode of frequency $\w_m$ will dominate over the energy carried by the primary mode $\hat\cs$ when $\hat\w/\Omega \lesssim v^{2\ell+6} \leq v^6$.

This phenomenon is particularly interesting if $\hat\cs$ is a counterrotating mode satisfying ${\hat m\Omega - \hat \w <0}$, since in this case we would predict an amplification factor ${Z < 0}$ when truncating to $\mathcal O(\mathcal A)$. However, if this ingoing mode has ${\hat\w \lesssim v^6 \Omega}$, the outgoing energy flux is dominated by the $\mathcal O(\mathcal A^2)$ term, which is positive definite; cf.~\eqref{eq:cont_scattering_2_dot_E}. Thus, we learn that energy is always extracted from the binary during this kind of scattering process if the ingoing frequency is low enough.

To be clear, while $\dot E_\text{SW}^{(2)}$ can be very large relative to $\dot E_\text{SW}^{(1)}$, its magnitude is still small in comparison to $\dot E_\text{SW}^\text{in}$ and further decreases as ${\hat\w \to 0}$, meaning the actual amount of energy that a low-frequency, counterrotating mode extracts from a binary is always negligible. Nonetheless, this calculation illustrates the kinds of rich physics that can arise as a consequence of mode mixing.

\section{Discussion}
\label{sec:discussion}

While we now have a comprehensive picture of how binary black holes evolve when in isolation, questions about their dynamical response to external perturbations are stimulating an emerging area of active research. The benefits to be reaped from this enterprise are twofold. First, the study of how ultralight fields influence the orbital evolution of a binary provides us with the prospect of using gravitational-wave detectors as tools to search for new physics. Second, even in the absence of a discovery, this kind of theoretical work offers new insight into general relativity and the properties of gravitational systems.

In this paper, we tracked the evolution of a long-wavelength scalar field living on a fixed binary black hole background. The interplay between absorption at the horizons and momentum transfer in the bulk gives rise to a novel energy-extraction mechanism, which we have herein dubbed ``orbital superradiance.'' It was previously shown that this mechanism can lead to the amplification of incident, low-frequency radiation \cite{Wong:2019kru}. The main novelty in this work is a nonperturbative treatment of the binary's long-range gravitational potential, which facilitates an extension of the results in Ref.~\cite{Wong:2019kru} to include the formation and evolution of bound states.

The key takeaway is as follows: Consider for simplicity an incident spherical wave that is peaked at some frequency $\hat\w$ and has angular momentum specified by the integers $(\hat\ell,\hat m)$. Three effects are triggered when this scalar wave scatters off a binary black hole. First, the ingoing mode is reflected back out with an amplitude that is either amplified or reduced. Amplification occurs if it corotates with the binary at an angular phase velocity $\hat\w/\hat m$ smaller than the binary's orbital frequency $\Omega$, while a net absorption of the mode occurs otherwise. Second, owing to the inherent lack of symmetries in this system,  multiple secondary modes are generated during scattering, which propagate outwards at certain frequencies given by ${ \w_m = |\hat \w + (m-\hat m)\Omega| }$. Third, a fraction of this ingoing wave is captured and converted into (quasi)bound states. It is worth noting that the binary's ``dumbbell'' geometry  establishes a selection rule that requires the integer ${\hat\ell+\hat m}$ to be even if any of these effects are to take place, since field configurations that violate this condition are unappreciable in the neighborhood of the binary. Likewise, only secondary outgoing modes and bound states with angular momenta satisfying ${\ell+m \in 2\mathbb Z}$ are generated during this scattering process.

The bound states that form around a binary subsequently evolve in an intricate manner due to a combination of three effects: First, orbital superradiance drives each bound state to either grow or decay exponentially, depending on its angular momentum. Second, the underlying geometry of the spacetime allows different modes to mix into one another, causing each bound state to exhibit a beating pattern by virtue of oscillating at multiple frequencies. For a scalar field of mass~$\mu$, the bound states with angular momentum $(\ell,m)$ oscillate at frequencies given approximately by ${ \w_{m'} \simeq |\mu + (m-m')\Omega| }$. Third, a fraction of these bound states are inevitably upscattered and ejected out of the system as scalar waves. (There is also a concomitant emission of gravitational waves from the cloud, although this is typically a subleading effect.) The rate at which outgoing scalar radiation depletes the energy in the scalar cloud can exceed the growth rates of the bound states when $\mu$ is considerably smaller than $\Omega$, in which case scalar clouds can no longer form dynamically around the binary. 

All of these effects illustrate the rich phenomenology that can arise in systems with horizons (or dissipative channels, more broadly) when time-translation invariance and rotational symmetry are weakly broken.%
\footnote{In the sense that both symmetries are restored in the EFT upon removal of the interaction terms, which we treated perturbatively.} 
Furthermore, the calculations underpinning these results demonstrate the usefulness of modern EFT methods in understanding the dynamics of complex systems with multiple hierarchies of scales. Unfortunately, they predict that orbital superradiance is grossly inefficient: the energy extracted from a binary black hole to amplify incident scalar radiation or to fuel the growth of bound states is always negligible for systems within the EFT's regime of validity. Moreover, the trends in Figs.~\ref{fig:growth} and \ref{fig:Z} suggest that this conclusion also extends to scalar fields with higher frequencies or larger masses, as long as the binary is in its early inspiral~phase.%
\footnote{Our perturbative approach breaks down when the binary is closer to merger, although the way the growth rates and amplification factors scale with the orbital velocity suggests that they may become appreciable in this regime. While this may be true, the binary does not remain in this stage for long and the usual version of black hole superradiance quickly takes over once the binary coalesces.}

While this is certainly disappointing from an observational standpoint, our results still constitute useful information about which effects play an important role during a binary's lifetime. Besides, orbital superradiance is expected to be just one of many phenomena that arise when a binary is perturbed by an external field. There is still much to do before a comprehensive survey of all of the effects that can occur is in~hand.

Natural next steps include relaxing some of the assumptions made in this paper. In particular, truncating to leading order in perturbation theory led us to neglect any interaction between the scalar field and the spins of the individual black holes. However, it has been shown that ambient matter generically exerts a ``gravitational Magnus force'' on spinning black holes \cite{Costa:2018gva}; hence, the presence of an external scalar field is likely to have an effect on the precession of the binary's orbital plane. Additionally, it is conceivable that generalizing to the case of eccentric orbits or including gravitational radiation from the binary will also teach us something new about how these systems interact with external fields. That being said, it is important to temper our expectations for observing any effect we study in this long-wavelength limit, since the large separations of scales inherent in this regime typically lead to strong power-law suppression by the EFT's expansion parameters.

Indeed, all work to date (including the upward trends in the right panels of Figs.~\ref{fig:growth} and \ref{fig:Z}) point to the likelihood of scalar fields with higher frequencies or larger masses having a more dramatic impact on the orbital evolution of a binary black hole; especially when resonant excitations can occur \cite{Bernard:2019nkv, Baumann:2019ztm}. Even so, studies in the long-wavelength limit will continue to be of value moving forward. The fact that we have a strong analytic handle on the problem in this regime can be used to gain physical intuition for better interpreting the results of numerical simulations, which may be the only recourse in certain scenarios. Of particular interest, for example, is the case of a scalar field whose characteristic size is comparable to the binary's orbital separation. The general ideas and  techniques found in this paper could also find applications in other branches of physics that involve open systems wherein one or more spacetime symmetries are weakly broken.

\acknowledgements
It is a pleasure to thank Vitor Cardoso, Anne-Christine Davis, Eugene Lim, and Ulrich Sperhake for stimulating discussions. This work was partially supported by STFC Consolidated Grants No.~ST/P000673/1 and No.~ST/P000681/1, and by a Cambridge Philosophical Society research studentship award (Ref.~S52/064/19). I am also supported in part by the Cambridge Commonwealth, European and International Trust and Trinity College, Cambridge.

\appendix
\section{Properties of the\protect\\radial solutions}
\label{sec:app_radial}
This Appendix provides a collection of useful identities for the radial solutions to~\eqref{eq:pert_eom_R}.

\paragraph{Limiting forms}
The identities in this first part have all been reproduced or adapted from Ref.~\cite{DLMF}. At large distances (${r\to\infty}$), the $R^\pm_\ell$ solutions have the asymptotic forms
\begin{align}
\label{eq:app_H_asymptotic}
	R^\pm_\ell(k,r) &= \frac{H_\ell^\pm(\zeta,kr)}{\pm ikr}
	\sim \frac{e^{\pm i \theta_\ell(\zeta,kr)} }{\pm ikr}[1 + \mathcal O(r^{-1})],
	\allowdisplaybreaks\\
	\theta_\ell(\zeta,kr) &= kr - \zeta \log(2kr) - \frac{\ell\pi}{2} + \sigma_\ell(\zeta).
\end{align}
If instead ${r \to 0}$, the behavior of the radial solutions around the origin may be inferred from the limiting forms of the Whittaker functions. The solutions that are regular at the origin are all proportional to
\bgroup\predisplaypenalty=0
\begin{equation}
	\rule{0pt}{15pt}
	M_{-i\zeta,\ell+1/2}(z) = z^{\ell+1}[1 + \mathcal O(z)],
\label{eq:app_M_limiting}
\end{equation}
\egroup
whereas the irregular solutions are proportional to
\begin{equation}
	W_{-i\zeta,\ell+1/2}(z)
	=
	\frac{\Gamma(2\ell+1)}{\Gamma(\ell+1+i\zeta)}
	\times
	\begin{cases}
		z^{-\ell}[ 1 + \mathcal O(z)] & (\ell\geq 1)
		\\
		1 + \mathcal O(z\log z) & (\ell=0).	
	\end{cases}
\end{equation}

It is also useful to know the limiting behavior of these solutions for small and large values of $\zeta$. For ${\zeta \to 0}$ with $k$ held fixed (corresponding to a removal of the gravitational potential), one has
\begin{equation}
	R^\pm_\ell(k,r) \sim h_\ell^\pm(kr),
	\quad
	R_\ell(k,r) \sim j_\ell(kr),
\end{equation}
where $h_\ell^\pm$ are the spherical Hankel functions while $j_\ell$ is the spherical Bessel function of the first kind. Instead taking the low-momentum limit ${k \to 0}$ (i.e., ${\zeta \to \infty}$  with $M$ and $\mu$ held fixed), one recovers the usual Bessel functions:
\begin{align}
	&
	\frac{(-\zeta)^\ell e^{\pi\zeta/2-i\sigma_\ell(\zeta)}}{\Gamma(\ell+1-i\zeta)}
	R_\ell(k,r)
	\sim
	\frac{ J_{2\ell+1}(2\sqrt{2GM\mu^2r}) }{\sqrt{2GM\mu^2r}},
	\\
	&
	\frac{ e^{\mp i\sigma_\ell(\zeta)} \Gamma(\ell+1\pm i\zeta) }%
	{\pm 2\pi i(-\zeta)^{\ell+1} e^{\pi\zeta/2}}
	R^\pm_\ell(k,r)
	\sim
	\frac{ Y_{2\ell+1}(2\sqrt{2GM\mu^2r}) }{\sqrt{2GM\mu^2r}}.
\label{eq:app_R+_limiting_Bessel}
\end{align}
Note that the prefactor multiplying $R^+_\ell$ on the lhs of \eqref{eq:app_R+_limiting_Bessel} has exactly the same $k$ dependence as the outgoing amplitude $\mathcal A^{(1)}_\cs$ in \eqref{eq:cont_A_main}, thus providing a good sanity check that the solution in \eqref{eq:cont_phi_1} is well behaved for all values of $k$.

\paragraph{Derivatives}
The result for the induced multipoles in \eqref{eq:pert_Olm_explicit} requires computing derivatives of the scalar field evaluated at the origin. Because $\partial_{L'}\phi(t,\bmf 0)$ is contracted with the STF product $\bmf z_N^\avg{L'}(t)$ in~\eqref{eq:eft_O}, only the STF part of the derivative contributes and note that we can write
\begin{equation}
	\partial_\avg{L}\phi(t,\bmf 0) \equiv \int\dx^3\bmf x\,\delta^{(3)}(\bmf x)\partial_\avg{L}\phi(t,\bmf x).
\end{equation}
Since both $R_\ell(k,r)$ and $R_{n\ell}(r)$ are proportional to the Whittaker function $M$, a good starting point is
\begin{align}
	&\int\dx^3\bmf x \, \delta^{(3)}(\bmf x) \partial_\avg{L'}
	\left(
		\frac{ M_{-i\zeta,\ell+1/2}(-2ikr) }{-2ikr} Y_{\ell m}(\hat{\bmf x})
	\right)
	\nonumber\\
	&=
	(-2ik)^\ell \ell! (\Y^{\ell m}_{L'}) \delta_{\ell\ell'},
\label{eq:app_derivative_Whittaker_M}
\end{align}
which follows from \eqref{eq:app_M_limiting} and the identity ${\partial_L \bmf x^L = \ell!}$. The definitions in \eqref{eq:app_R_analytic_cont} and \eqref{eq:app_Rnl_Whittaker} can then be used to show that
\begin{subequations}
\label{eq:app_derivatives_mode_functions}
\begin{align}
	\frac{4\pi}{(2\ell'+1)!!} (\Y^{\ell'm'}_{L'})^*
	\partial_{L'} \phi_{k\ell m}(\bmf 0)
	&=
	\mathfrak R_\ell(k) \delta^{\ell\ell'}\delta^{mm'},
	\\
	\frac{4\pi}{(2\ell'+1)!!} (\Y^{\ell'm'}_{L'})^*
	\partial_{L'} \psi_{n\ell m}(\bmf 0)
	&=
	\mathfrak R_{n\ell} \delta^{\ell\ell'}\delta^{mm'},
\end{align}
\end{subequations}
where the coefficients on the rhs are given by
\begin{subequations}
\label{eq:app_derivative_R}
\begin{align}
	\mathfrak R_\ell(k)
	&=
	C_\ell(\zeta) k^\ell,
	\\
	\mathfrak R_{n\ell}
	&=
	\sqrt\frac{(n+\ell)!}{2(n-\ell-1)!}
	\frac{ (2GM\mu^2)^{\ell+3/2} }{ (2\ell+1)! n^{\ell+2} }.
\end{align}
\end{subequations}

\paragraph{Complex conjugates}
Several instances in the main text exploit identities for the complex conjugates of the mode functions to obtain simplified expressions. These identities are derived here. Combining the well-known identity
\begin{equation}
	Y_{\ell m}^*(\hat{\bmf x}) \equiv (-1)^m Y_{\ell,-m}(\hat{\bmf x})
\label{eq:app_cc_Y}
\end{equation}
with the fact that $R_{n\ell}(r)$ is a real function tells us that the complex conjugate of a bound-state mode function is
\begin{equation}
	\psi_{n\ell m}^*(\bmf x) \equiv (-1)^m\psi_{n\ell-m}(\bmf x).
\label{eq:app_cc_psi}
\end{equation}
As for the continuum states, the identity~\cite{Gaspard:2018xgb}
\begin{equation}
	[H_\ell^\pm(\zeta,z)]^* \equiv H^\mp_\ell(\zeta^*,z^*)
\end{equation}
can be used to show that
\begin{equation}
	[R^\pm_\ell(k,r)]^* \equiv R^\mp_\ell(k^*,r),
	\quad
	[R_\ell(k,r)]^* \equiv R_\ell(k^*,r).
\end{equation}
These can be written in a more useful form by utilizing the circuital relations~\cite{Gaspard:2018xgb,Dzieciol:1999}
\begin{subequations}
\begin{align}
	F_\ell(\zeta,z) &\equiv -e^{i\pi(\ell + i\zeta)} F_\ell(-\zeta, z e^{-i\pi}),
	\\
	H_\ell^+(\zeta,z) &\equiv e^{-i\pi(\ell + i\zeta)} H_\ell^-(-\zeta,z e^{-i\pi})
\end{align}
\end{subequations}
in conjunction with the identity ${k^*(\w) \equiv -k(-\w)}$, which is a consequence of the definition in \eqref{eq:pert_k}. Only the results for $R^+_\ell$ and $R_\ell$ are relevant for physical solutions. They are
\begin{subequations}
\begin{align}
	[R^+_\ell(k,r)]^* &\equiv e^{+i\pi(\ell + i\bar\zeta)} R^+_\ell(\bar k, r),
	\\[-0.1em]
	[R_\ell(k,r)]^* &\equiv e^{-i\pi(\ell + i\bar\zeta)} R_\ell(\bar k, r),
\end{align}
\end{subequations}
where we write ${\bar k \equiv k(-\w)}$ and ${\bar\zeta \equiv \zeta(-\w)}$ as shorthand. Combined with \eqref{eq:app_cc_Y}, the complex conjugates of the continuum-state mode functions are
\begin{subequations}
\begin{align}
	\phi^{+*}_{k\ell m}(\bmf x)
	&=
	e^{-\pi \bar\zeta} (-1)^{\ell+m} \phi^+_{\bar k \ell -m}(\bmf x),
	\label{eq:app_cc_phi_+}
	\\[-0.1em]
	\phi^*_{k\ell m}(\bmf x)
	&=
	e^{+\pi\bar\zeta} (-1)^{\ell+m} \phi_{\bar k \ell -m}(\bmf x).
	\label{eq:app_cc_phi_0}
\end{align}
\end{subequations}

\section{Resummation and late-time\protect\\behavior of bound states}
\label{sec:app_resum}

The growth of $\Gamma_\bs t$ in \eqref{eq:bound_linear_growth} invalidates our naive perturbative approach once it becomes of order unity, even though the expansion parameters enumerated in Sec.~\ref{sec:eft} all remain small. This kind of secular growth turns out to be generic in any system with an interaction Hamiltonian that persists for all times \cite{Burgess:2018sou}. For the scenario studied in this paper, this late-time breakdown of perturbation theory poses no threat because the binary will typically coalesce well before ${\Gamma_\bs t \sim 1}$.  That being said, on theoretical grounds, it is interesting to explore how we might obtain an approximate solution to \eqref{eq:eft_eom} that remains valid at late times. The general results may find application in studies of other open systems whose lifetimes exceed $1/\Gamma_\bs$.

The key is to carefully resum the dominant polynomial behavior ${\propto t^p}$ at each order $p$ in perturbation theory while neglecting subleading terms. Included in this set of terms we will neglect are backreaction effects from the outgoing radiation and Yukawa modes (see Sec.~\ref{sec:cont}), because they contribute to $c_\bs(t)$ beginning only at second order. Additionally, higher-order corrections to the formula for the induced multipoles in \eqref{eq:eft_O}, which are suppressed by extra powers of $v$ and $GM\w$, can also be neglected. For added simplicity, we will also assume no ingoing radiation in this Appendix.

With these considerations in mind, resummation amounts to looking for a solution to \eqref{eq:eft_eom} of the form ${\phi(x) = \sum_{p=0}^\infty\phi^{(p)}(x)}$, where each term in this series is sourced by the previous term via the iteration
\begin{align}
	\phi^{(p+1)}(x)
	&=
	\int\dx^4x' G(x,x')\sum_{\ell=0}^\infty (-1)^\ell O^{(p)}_L(t') 
	\partial_L \delta^{(3)}(\bmf x')
	\nonumber\\
	&\quad+
	\phi^{(p)}_\text{cf}(x).
\label{eq:pert_phi_p}
\end{align}
This is, of course, simply a generalization of \eqref{eq:pert_phi_1}. Accordingly, the coefficients for the bound states at order ${p+1}$ are given by
\begin{align}
	c^{(p+1)}_\bs(t)
	&=
	\sum_{\bs'} \frac{ V_{\bs\bs'} }{2E_n} \int^t\dx t'
	\big[
		(m'\Omega - E_{n'}) c^{(p)}_{\bs'} -i \dot c^{(p)}_{\bs'}
	\big]
	\nonumber\\&\quad\times
	\big( e^{i\Delta_{\bs\bs'}t'} - e^{2i E_n t} e^{i(\Delta_{\bs\bs'} - 2 E_n)t'} \big)
\end{align}
after suitably generalizing \eqref{eq:bound_c1_exp}. Rather than perform this string of integrals, the trick is to now differentiate twice to obtain
\begin{align}
	&\ddot c_\bs^{(p+1)}
	-
	2i E_n \dot c_\bs^{(p+1)}
	\nonumber\\
	&=
	-i\sum_{\bs'} V_{\bs\bs'}
	[ (m'\Omega - E_{n'}) c^{(p)}_{\bs'} -i \dot c^{(p)}_{\bs'} ]
	e^{i\Delta_{\bs\bs'}t}.
\label{eq:bound_resum_ODE_iterative}
\end{align}
In the same way that \eqref{eq:pert_phi_p} is an iterative solution to the equation of motion in \eqref{eq:eft_eom}, this set of differential equations in \eqref{eq:bound_resum_ODE_iterative} can be viewed as establishing an iterative method (assuming $V$ is suitably small) for solving the master equation%
\footnote{With the benefit of hindsight, this master equation can be seen to follow more easily from substituting the ansatz $\phi(x) \propto \sum_\bs[ c_\bs(t) \psi_\bs(\bmf x) e^{-i E_n t} + \text{c.c.}]$ directly into \eqref{eq:eft_eom}. However, doing so obscures the fact that the bound states alone are not a complete solution to the problem. As we discussed in Sec.~\ref{sec:cont_ejection}, the production of outgoing radiation is inevitable in this system.}%
\begin{equation}
	\ddot c_\bs - 2i E_n \dot c_\bs
	=
	-i\sum_{\bs'} V_{\bs\bs'}
	[(m'\Omega - E_{n'}) c_{\bs'} -i \dot c_{\bs'}] e^{i\Delta_{\bs\bs'} t}.
\label{eq:bound_resum_ODE_master}
\end{equation}

This equation is strongly reminiscent of time-dependent perturbation theory in quantum mechanics, albeit with two key differences. First, the terms on the rhs can be regarded as arising from some interaction Hamiltonian for the system, whereby
$\langle\bs | H_\text{int} | \bs' \rangle \propto -i V_{\bs\bs'} e^{-i(m-m')\Omega t} $.
Given that the diagonal elements $V_{\bs\bs}$ are real, the prefactor of $-i$ indicates that $H_\text{int}$ is not Hermitian---a necessary condition for this system to exhibit nonunitary evolution. The other key difference is that \eqref{eq:bound_resum_ODE_master} is clearly a set of second-order, rather than first-order, differential equations; reflecting the relativistic nature of~this~system.

To perform the requisite resummation, we now treat the terms involving the diagonal elements $V_{\bs\bs}$ on the rhs of \eqref{eq:bound_resum_ODE_master} nonperturbatively. Moving them over to the lhs, the master equation may be rewritten as
\begin{equation}
	\ddot c_\bs - (2i E_n- V_{\bs\bs}) \dot c_\bs
	+ 2i E_n \Gamma_\bs c_\bs
	=
	-i J_\bs,
\SetDisplaySkip[][0.4]
\end{equation}
where
\begin{equation}
	J_\bs(t)
	=
	\sum_{\bs'\neq\bs} V_{\bs\bs'}
	[(m'\Omega - E_{n'}) c_{\bs'} -i \dot c_{\bs'}] e^{i\Delta_{\bs\bs'} t}
\end{equation}
is independent of $c_\bs$ and can therefore be regarded as a source term. To solve this equation, we begin by noting that it is of the form
\begin{equation}
\SetDisplaySkip[0.3]
	\ddot c_\bs -(\gamma_+ + \gamma_-) \dot c_\bs + \gamma_+\gamma_- c_\bs = -i J_\bs,
\label{eq:bound_resum_ODE_simple}
\end{equation}
where $\gamma_\pm$ are the two zeros of the characteristic polynomial
${ \gamma_\pm^2 - (2i E_n - V_{\bs\bs}) \gamma_\pm + 2i E_n \Gamma_\bs}$. As ${V \ll E_n}$, it suffices to use the approximate solutions
\begin{subequations}
\begin{align}
	\gamma_+ &\simeq 2 iE_n - (\Gamma_\bs + V_{\bs\bs}) = 2 iE_n + \Gamma_{\bar\bs},
	\\
	\gamma_- &\simeq \Gamma_\bs,
\end{align}
\end{subequations}
where we define ${\bar\bs \equiv (n,\ell,-m)}$ as shorthand. For later purposes, it will also be useful to define $\gamma_\bs \coloneq (\gamma_+ - \gamma_-)/2$. Now choosing boundary conditions such that $c_\bs(0) = c^{(0)}_\bs$ and $c_\bs(t) \to c^{(0)}_\bs \;\forall\,t$ in the limit ${V \to 0}$, the solution to \eqref{eq:bound_resum_ODE_simple} is
\begin{equation}
	c_\bs(t) = 
	c^{(0)}_\bs e^{\gamma_- t}
	+
	\frac{i}{2\gamma_\bs} \int_0^t\dx t'
	( e^{\gamma_- (t-t')} - e^{\gamma_+ (t-t')} )  J_\bs(t').
\label{eq:bound_resum_sol_formal}
\end{equation}

This is only a formal solution because $J_\bs$ depends on the other bound states ${\bs' \neq \bs}$, whose solutions are also given by \eqref{eq:bound_resum_sol_formal}. To obtain an explicit result, we iterate \eqref{eq:bound_resum_sol_formal} in powers of ${{V/\gamma} \ll 1}$. Starting with
${ c_\bs(t) = c^{(0)}_\bs e^{\Gamma_\bs t} + \mathcal O(V/\gamma) }$,
after one iteration we find
\bgroup\AddDisplaySkip[-0.3][-0.8]
\begin{align}
	c_\bs(t)
	&=
	c^{(0)}_\bs e^{\Gamma_\bs t}
	+
	\sum_{\bs'\neq\bs} \frac{ V_{\bs\bs'} }{2\gamma_\bs}
	(m'\Omega - \tilde E_{\bs'})
	\bigg[
		\frac{ e^{i \Delta_{\bs\bs'}t} e^{\Gamma_{\bs'}t}
		- e^{\Gamma_\bs t} }{\tilde\Delta_{\bs\bs'} }
		\nonumber\\&\quad
		-
		\frac{ e^{i\Delta_{\bs\bs'}t} e^{\Gamma_{\bs'}t}
		- e^{2i E_n t} e^{\Gamma_{\bar\bs} t} }%
		{\tilde\Delta_{\bs\bs'} + 2i\gamma_\bs}
	\bigg]c_{\bs'}^{(0)}
	+
	\mathcal O( (V/\gamma)^2 ),
\label{eq:bound_resum_sol_final}
\end{align}
\egroup
where ${\tilde E_\bs = E_n + i\Gamma_\bs}$ is the complex frequency of the quasibound state and $\tilde\Delta_{\bs\bs'}$ is defined in the same way as $\Delta_{\bs\bs'}$ in \eqref{eq:bound_Delta_bb} except with $\tilde E_\bs$ in place of $E_\bs$. Having carefully resummed the leading polynomial growth to all orders, this solution is valid at late times ${t \gg 1/\Gamma_\bs}$ while still being organized as a perturbative expansion in the small parameter $V/\gamma$.

This resummed solution also brings with it a new prediction. Let us denote the fastest-growing mode by $\bs_\star$ and its growth rate by $\Gamma_\star$. The sum over $\bs'$ in \eqref{eq:bound_resum_sol_final}, which quantifies the leading effects due to mode mixing, then tells us that all modes satisfying ${V_{\bs \bs_\star} \neq 0}$ will grow at the same rate $\Gamma_\star$ at late times, even if they were initially decaying.%
\footnote{This phenomenon occurs when the $\mathcal O(V/\gamma)$ terms dominate over the first term in \eqref{eq:bound_resum_sol_final}. Nonetheless, our perturbative expansion is still valid because the $\mathcal O((V/\gamma)^2)$ terms remain subleading.} 
While this is a nontrivial result for the system of equations under study, it is irrelevant in the case of a scalar cloud around a binary black hole because even the shortest $e$-folding time $1/\Gamma_\star$ is always orders of magnitude greater than the orbital decay timescale $1/\Gamma_\text{GW}$. It would therefore be interesting to explore if there are other open systems that could survive long enough to exhibit this universal growth rate at late times.

\section{Scalar-wave flux}
\label{sec:app_flux}

Multiple instances in Sec.~\ref{sec:cont} call for the evaluation of the power radiated to infinity in scalar waves. For the sake of efficiency, we derive here a general formula for the energy flux when it is sourced by a single bound state $\hat\bs$ or a single ingoing radiation mode $\hat\cs$. In either of these cases, the components of the induced multipoles take on the general form
\begin{equation}
	O^{(0)}_{\ell m}(t) = o_{\ell m} e^{-i \w_m t} - (-1)^m o^*_{\ell,-m} e^{i \w_{-m} t}.
\label{eq:app_energy_Olm}
\end{equation}
When compared to \eqref{eq:pert_Olm_explicit}, we see that the complex coefficients $o_{\ell m}$ and the real frequencies $\w_m$ are given by
\bgroup\predisplaypenalty=0
\begin{subequations}
\begin{align}
	o_{\ell m}
	&=
	\frac{1}{\sqrt{2\mu}}
	\frac{Y^*_{\ell m}(\bmf d) Y_{\hat\ell\hat m}(\bmf d) B_{\ell\hat\ell}}{(2\ell+1)!!}
	\mathfrak R_{\hat n\hat\ell} (\hat m\Omega - E_{\hat n}) c_{\hat\bs}^{(0)},
	\\
	\w_m
	&=
	E_{\hat n} + (m-\hat m)\Omega
\end{align}
\end{subequations}
\egroup
for the case of a single bound state ${\hat\bs \equiv (\hat n,\hat\ell, \hat m)}$. If instead we considered a single ingoing radiation mode ${\hat\cs \equiv (\hat\w,\hat\ell,\hat m)}$ with $\mathcal I^>_\cs$ given by \eqref{eq:cont_single_ingoing_rad_mode}, we would have
\begin{subequations}
\begin{align}
	o_{\ell m}
	&=
	\frac{Y^*_{\ell m}(\bmf d) Y_{\hat\ell\hat m}(\bmf d) B_{\ell\hat\ell}}{(2\ell+1)!!}
	\mathfrak R_{\hat\ell}(\hat k) (\hat m\Omega - \hat\w) 2\Phi_{\hat\cs}
	\\
	\w_m
	&=
	\hat\w + (m-\hat m)\Omega.
\end{align}
\end{subequations}

Substituting the general form for $O_{\ell m}^{(0)}(t)$ in \eqref{eq:app_energy_Olm} into \eqref{eq:cont_A}, the first-order correction to the outgoing amplitude reads
\begin{align}
	\mathcal A^{(1)}_\cs &=
	s_\ell(\zeta) k^{\ell+1}
	\big[
		2\pi\delta(\w-\w_m) o_{\ell m}
		\nonumber\\&\quad
		- (-1)^m 2\pi\delta(\w + \w_{-m}) o^*_{\ell,-m}
	\big].
\label{eq:app_energy_A1}
\end{align} 
Because the energy flux is quadratic in the scalar field [cf.~\eqref{eq:cont_dot_E}], there are contributions at first and second order in $\mathcal A^{(1)}_\cs$. Written out explicitly, the total energy lost to scalar radiation is
\begin{equation}
	\int\dx t\, \dot E_\text{SW} = \sum_\cs \theta(k^2) \frac{\w}{k}
	\big( 2\Re \mathcal I^*_\cs \mathcal A^{(1)}_\cs + |\mathcal A^{(1)}_\cs|^2 \big).
\label{eq:app_energy_dotE_rad}	
\end{equation}

\paragraph{First order}
Let us begin by evaluating the term in \eqref{eq:app_energy_dotE_rad} that is linear in $\mathcal A^{(1)}_\cs$. Using \eqref{eq:app_energy_A1}, the power loss at first order is
\begin{align}
	\int\dx t\,\dot E_\text{SW}^{(1)}
	&=
	2 \Re \sum_\cs \theta(k^2) s_\ell(\zeta)
	\big[
		2\pi\delta(\w-\w_m) o_{\ell m}
		\nonumber\\&\quad
		- (-1)^m 2\pi\delta(\w + \w_{-m}) o^*_{\ell,-m}
	\big]
	\mathcal I^*_\cs \w k^\ell.
\end{align}
To proceed, we deduce from the definition of $s_\ell(\zeta)$ in \eqref{eq:cont_s} that $s_\ell(-\zeta) = s_\ell^*(\zeta) e^{\pi\zeta}$ when ${\zeta\in\mathbb R}$ (${k^2>0}$). Combined with the freedom to relabel ${\w \to -\w}$ and ${m\to -m}$ as they are being integrated and summed over, respectively, we find
\begin{align}
	\int\dx t\,\dot E_\text{SW}^{(1)}
	&=
	2 \Re \sum_\cs \theta(k^2)
	2\pi\delta(\w - \w_m)
	\big[
		\mathcal I^*_{\w\ell m} s_\ell(\zeta) o_{\ell m}
		\nonumber\\&\quad
		+
		e^{\pi\zeta}(-1)^{\ell+m} \mathcal I^*_{-\w\ell-m}
		s_\ell^*(\zeta) o^*_{\ell m} 	
	\big] \w k^\ell.
\end{align}
The identity in \eqref{eq:cont_I_constraint} may now be used to show that the terms in square brackets are complex conjugates of one another, and thus this expression further simplifies to
\begin{equation}
	\int\dx t\,\dot E_\text{SW}^{(1)}
	=
	4 \Re \sum_\cs \theta(k^2)
	2\pi\delta(\w - \w_m) o_{\ell m}
	s_\ell(\zeta) \w k^\ell \mathcal I^*_\cs.
\label{eq:app_energy_P_1}
\end{equation}

Of course, this general result is valid only for induced multipoles sourced by a single bound state $\hat\bs$ or a single ingoing radiation mode $\hat\cs$. If the former, we would have $\mathcal I^*_\cs = 0 \;\forall\,\cs$, meaning $\mathcal A^{(1)}_\cs$ contributes linearly to the energy flux only if it interferes with ingoing radiation. Performing the sum over $\cs$ in \eqref{eq:app_energy_P_1}, we pick up a single nonvanishing contribution at the frequency ${\w_{\hat m}\equiv \hat\w}$, which yields
\begin{equation}
	\dot E_\text{SW}^{(1)}
	=
	4\Re s_{\hat\ell}(\hat\zeta) \hat\w\hat k^{\hat\ell}
	o_{\hat\ell\hat m} \Phi_{\hat\cs}^*.
\label{eq:app_energy_P_1_final}
\end{equation}
In obtaining this result, notice that the delta function $2\pi\delta(0)$ contained implicitly in $\mathcal I^*_{\hat\w\hat\ell\hat m}$ cancels against the integral over all time, $\int\dx t\equiv 2\pi\delta(0)$.

\paragraph{Second order}
In the absence of ingoing radiation, the leading contribution to the energy flux is quadratic in $\mathcal A^{(1)}_\cs$. Taking the absolute square of \eqref{eq:app_energy_A1}, one finds
\begin{align}
	\big| \mathcal A_w^{(1)} \big|^2
	&=
	S_\ell(\zeta) k^{2(\ell+1)}
	(2\pi)^2
	\big[
		\delta(\w-\w_m)\delta(0)
		|o_{\ell m}|^2
		\nonumber\\&\quad
		+
		\delta(\w + \w_{-m})\delta(0)
		|o_{\ell,-m}|^2
	\big].
\label{eq:app_energy_A2_explicit}
\end{align}
To arrive at this result, we use the fact that the cross terms proportional to $\delta(\w - \w_m)\delta(\w + \w_{-m})$ may be discarded because they have nonoverlapping support.%
\footnote{The frequencies generally satisfy the condition $\w_{m} \neq - \w_{-m}$ except when ${\hat m\Omega - E_{\hat n} = 0}$ or ${\hat m\Omega - \hat\w = 0}$, in which case the coefficient $o_{\ell m}$ vanishes.}
It then follows that
\begin{align}
	\dot E_\text{SW}^{(2)}
	&=
	\frac{1}{2\pi\delta(0)} \sum_\cs \theta(k^2) \frac{\w}{k}
	\big| \mathcal A^{(1)}_\cs \big|^2
	\allowdisplaybreaks\\
	&=
	\sum_\cs \theta(k^2)
	S_\ell(\zeta) \w k^{2\ell+1} |o_{\ell m}|^2
	\nonumber\\&\quad\times
	[ 2\pi\delta(\w-\w_m) +2\pi\delta(\w + \w_m)],
\end{align}
after also using the freedom to relabel ${m \to -m}$. To simplify this result one step further, we note that the product $\w k^{2\ell+1}$ is invariant under the transformation ${\w \to -\w}$, whereas $\zeta$ changes sign; hence, we may easily perform the integral over $\w$ to obtain
\begin{equation}
	\dot E_\text{SW}^{(2)}
	=
	\sum_{\ell,m} \theta(k_m^2)
	[S_\ell(\zeta_m) + S_\ell(-\zeta_m)] |o_{\ell m}|^2 \w_m k_m^{2\ell+1},
\end{equation}
where $k_m \equiv k(\w_m)$ and likewise $\zeta_m\equiv\zeta(\w_m)$.

\paragraph{Ingoing flux}
The ratio of $\dot E_\text{SW}$ to $\dot E_\text{SW}^\text{in}$ is often a useful measure. For a single ingoing mode $\hat\cs$, the latter is given by substituting \eqref{eq:cont_single_ingoing_rad_mode} into \eqref{eq:cont_dot_E_in}. After neglecting cross terms that involve products of delta functions with nonoverlapping support, the end result is
\begin{equation}
	\dot E_\text{SW}^\text{in}
	=
	\frac{\hat\w}{\hat k}|\Phi_{\hat\cs}|^2 (1 + e^{-2\pi\hat\zeta}),
\label{eq:app_energy_ingoing_flux}
\end{equation}
where the exponential arises from the identity in \eqref{eq:cont_I_constraint}.

\section{Different mass ratios}
\label{sec:app_mass_ratio}

For a binary composed of spherical (sph) black holes, the growth rate of the ${\bs \equiv (n,\ell,m)}$ mode may be written as
\begin{align}
	\left( \frac{\Gamma_\bs}{\Gamma_\text{GW}} \right)_\text{sph}
	&=
	\frac{\mathfrak n_\ell(\nu)}{\nu}
	\left| \frac{Y_{\ell m}(\bmf d)}{(2\ell+1)!}\right|^2
	\frac{5\pi  (n+\ell)! }{ (n-\ell-1)! 4n^{2\ell+4} } 
	\nonumber\\&\quad\times
	v^{8\ell+10} f_\mu^{4\ell+5} (m - f_\mu)
\label{eq:app_mr_Gamma}
\end{align}
in terms of the dimensionless parameters $v$, ${\nu \coloneq M_1 M_2/M^2}$, and ${f_\mu\coloneq\mu/\Omega}$. Similarly, the amplification factor for the $\cs \equiv (\w,\ell,m)$ mode is
\begin{align}
	Z_\text{sph} &=
	\mathfrak n_\ell(\nu)
	\bigg( \frac{2S_{\ell}(\zeta)}{1 + e^{-2\pi\zeta}} \bigg)
	\left| \frac{Y_{\ell m}(\bmf d)}{(2\ell+1)!!}\right|^2
	\frac{32\pi}{ 4^\ell}
	\nonumber\\&\quad\times
	v^{2\ell+6}
	(f_\w^2 - f_\mu^2)^{\ell+1/2}
	(m-f_\w).
\label{eq:app_mr_Z}
\end{align}
In this case, the expression depends on the four dimensionless quantities $v$, $\nu$, $f_\mu$, and $f_\w \coloneq \w/\Omega$, and we note that $\zeta \equiv - v^3 {(f_\w^2/f_\mu^2 - 1)^{-1/2}}$. In both formulas, the effect of the symmetric mass ratio $\nu$ enters via the same function
\begin{equation}
	\mathfrak n_\ell(\nu) \coloneq 8\nu^2
	\big[ (1+\sqrt{1-4\nu})^{2\ell-2} + (1-\sqrt{1-4\nu})^{2\ell-2} \big],
\end{equation}
which is normalized such that ${ \mathfrak n_\ell(1/4)=1 }$.

The additional prefactor of $1/\nu$ in \eqref{eq:app_mr_Gamma} causes the ratio $\Gamma_\bs/\Gamma_\text{GW}$ to diverge in the limit ${\nu \to 0}$. This singularity is unrelated to $\Gamma_\bs$ and is due entirely to $\Gamma_\text{GW}$. Physically, it is reflecting the fact that the timescale over which the orbit shrinks becomes infinite in the limit of a test particle around a host black hole. It follows that $\Gamma_\bs$ and $Z$ are both proportional to $\mathfrak n_\ell(\nu)$; hence, our discussion will focus purely on this function's properties.

It presents three different classes of behavior depending on the value of~$\ell$. When $\ell=0$, the largest value of $\mathfrak n_0(\nu) = 2-4\nu$ in the domain $\nu \in (0,1/4]$ coincides with $\nu=0$. This behavior has a simple physical interpretation: for a binary with fixed total mass $M$, a smaller symmetric mass ratio leads to a larger combined area for the black hole's horizons. In other words, that $\mathfrak n_0(\nu)$ is maximized when ${\nu=0}$ simple corroborates the fact that absorption is more efficient when there is a larger horizon area. Note, however, that the value of this function only changes by a factor of $2$ in the domain $\nu \in (0,1/4]$.

When ${\ell=1}$, one finds ${\mathfrak n_1(\nu) = 16\nu^2}$, which is maximized when the binary's components have equal masses (${\nu=1/4}$). For ${\ell \geq 2}$, this function always has a maximum somewhere in the domain ${\nu \in (0,1/4]}$. Numerically, we find
$\max\mathfrak n_\ell(\nu) \sim \exp(1.4 \ell -2.0\log\ell - 1.5)$ when ${\ell \gg 1}$, which can be quite a large number. For instance, $\max \mathfrak n_{20}(\nu) \sim 10^9$.

What does this mean for the conclusions in the main text? Given that $\mathfrak n_1(\nu)$ is maximized for equal-mass binaries, the growth rates for the ${\ell=1}$ modes (shown in Fig.~\ref{fig:growth}) and the amplification factors for the same modes (shown in Fig.~\ref{fig:Z}) are indeed the largest values possible within the EFT's regime of validity. For larger values of $\ell$, carefully selecting an optimal value for $\nu$ can enhance the growth rates and amplification factors relative to the equal-mass case, but this enhancement grows exponentially with $\ell$ at best, which is still no match for the factorials in the denominators of \eqref{eq:app_mr_Gamma} and \eqref{eq:app_mr_Z}. Consequently, the general trend remains unchanged: the maximum value that $\Gamma_\bs$ or $Z$ can attain decreases rapidly as we increase $\mu$ or $\w$.

\newpage

\bibliography{cloud}
\end{document}